\documentstyle[11pt,epsf]{article}

\newcommand{\be}{\begin{equation}}
\newcommand{\ee}{\end{equation}}
\newcommand{\nn}{\nonumber}
\newcommand{\beba}{\begin{equation}\begin{array}{lcl}}
\newcommand{\eaee}{\end{array}\end{equation}}
\newcommand{\bea}{\begin{eqnarray}}
\newcommand{\eea}{\end{eqnarray}}
\newcommand{\ba}{\begin{array}}
\newcommand{\ea}{\end{array}}

\newcommand{\ns}{\normalsize}
\newcommand{\refs}[1]{(\ref{#1})}
\def\bal{{\mbox{\boldmath $\alpha$}}}
\def\bla{{\mbox{\boldmath $\lambda$}}}
\def\bbe{{\mbox{\boldmath $\beta$}}}
\def\bt{{\mbox{\boldmath $\tau$}}}
\def\bq{{\bf q}}
\def\bd{{\bf d}}
\def\bk{{\bf k}}
\def\bc{{\bf c}}
\def\bw{{\bf w}}
\def\bH{{\bf H}}
\def\bp{{\bf p}}
\def\bk{{\bf k}}
\def\bx{{\bf x}}
\def\boe{{\bf e}}

\def\a{\alpha}

\def\d{\delta}
\def\e{\epsilon}

\def\f{\phi}

\def\m{\mu}
\def\n{\nu}

\def\r{\rho}

\def\t{\tau}

\def\D{\Delta}

\def\L{\Lambda}
\def\O{\Omega}

\begin{document}


\begin{titlepage}
\title{\hfill{\ns UPR-723T, IASSNS-HEP-96/107, PUPT-1656\\}
       \hfill{\ns hep-th/9610238\\[.1cm]}
       \hfill{\ns October 1996}\\[.8cm]
       {\large\bf String and M--Theory Cosmological Solutions
              with Ramond Forms}}
\author{Andr\'e
        Lukas$^1$\setcounter{footnote}{0}\thanks{Supported by Deutsche
        Forschungsgemeinschaft (DFG) and
        Nato Collaborative Research Grant CRG.~940784.}~~,
        Burt A.~Ovrut$^1\; ^3$ and Daniel Waldram$^2$\\[0.5cm]
        {\ns $^1$Department of Physics, University of Pennsylvania} \\
        {\ns Philadelphia, PA 19104--6396, USA}\\[0.3cm]
        {\ns $^2$Department of Physics}\\
        {\ns Joseph Henry Laboratories, Princeton University}\\
        {\ns Princeton, NJ 08544, USA}\\[0.3cm]
        {\ns $^3$School of Natural Sciences, Institute for Advanced Study}\\
        {\ns Olden Lane, Princeton, NJ 08540, USA}}
\date{}
\maketitle

\begin{abstract} 
\thispagestyle{empty}

A general framework for studying a large class of cosmological solutions of 
the low-energy limit of type II string theory and of M--theory, with
non-trivial Ramond form fields excited, is presented. The framework is
applicable to spacetimes decomposable into a set of flat or, more generally,
maximally symmetric spatial subspaces, with multiple non-trivial form fields
spanning one or more of the subspaces. It is shown that the corresponding
low-energy equations of motion are equivalent to those describing a
particle moving in a moduli space consisting of the scale factors of
the subspaces together with the dilaton. The choice of which form
fields are excited controls the potential term in the particle
equations. Two classes of exact solutions are given, those
corresponding to exciting only a single form and those with multiple
forms excited which correspond to Toda theories.
Although typically these solutions begin or end in a curvature
singularity, there is a subclass with positive spatial curvature which
appears to be singularity free. Elements of this class are directly
related to certain black $p$-brane solutions. 

\end{abstract}

\thispagestyle{empty}
\end{titlepage}


\section{Introduction}

An important constraint on string theory or any generalization of
string theory, such as M--theory, is that it should be compatible with
the standard model of early universe cosmology. The standard approach to
string cosmology has been to consider an epoch when the evolution of the
universe is described by the string low-energy effective action. One
encounters a number of the familiar problems in matching this
description to standard cosmology~\cite{bru_stei}. In particular, it is
necessary to invoke some mechanism for stabilizing the massless string
moduli such as the dilaton. While understanding these issues may in fact
be central to understanding string cosmology \cite{mod_cos}, the
usual approach has been at least to start by studying the simple
cosmological solutions of the pure effective supergravity theory. 

In the past, most focus has been on heterotic strings as the best
model of low-energy particle physics. Consequently, study of
cosmological solutions has
concentrated on the dynamics of the dilaton and the compactified space
in cosmological solutions of the effective theory. However, with the
discovery of string dualities \cite{string_dual} and the
existence of D--brane states \cite{D-brane}, the
nature of string theory has changed dramatically. Strong-weak coupling
duality symmetries connect each of the five consistent supersymmetric
string theories together with eleven-dimensional supergravity
\cite{duff_rep}. As a consequence, type II and eleven-dimensional
supergravities may now be directly relevant to low-energy particle physics
and cosmology \cite{banks_dine}. Both theories contain form fields,
namely a three form in eleven-dimensions and Ramond-Ramond (RR) forms of
various degrees in type II theories. The importance of considering
the excitation of these fields has been stressed by the
discovery of objects carrying RR form-field charge, the D--branes of open
string theory, and the central role these states have played in
understanding the spacetime structure of black holes
\cite{black_hole}. Given this change of perspective, it
clearly becomes important to study the cosmological solutions of type
II and eleven dimensional supergravity with non-trivial form fields
excited. In a recent letter \cite{letter}, we gave the first study of a
class of such solutions, giving specific examples with one or more
forms excited. The purpose of this paper is to provide a more detailed
and complete description of this class of solutions.

The gravitational part of our Ansatz is based on the standard 
``rolling radii'' solutions of Mueller \cite{mueller}, which are a
generalization the classical Kasner solutions of Kaluza-Klein gravity
\cite{kasner}. In the original Mueller solutions spacetime is divided
into a set of 
maximally symmetric spatial subspaces, each with its own time-dependent
scale factor. (One subspace should be three-dimensional to describe our
observed spacetime.) The rate of expansion or contraction of the subspaces
is controlled by a power law and, typically, the universe either starts or
ends in a curvature singularity. Including a dilaton in ten dimensions does not
significantly alter the solution. This is not surprising given that one may
consider the dilaton as the radius of a compact eleventh dimension.
As was first pointed out in \cite{gibb_town}, in general such an 
evolution describes a motion on a moduli space with coordinates given
by the values of the scale factors together with the dilaton. (We
should point out that this is not quite the conventional notion of a
string moduli space. For instance, it includes a scale factor
``modulus'' for our observed four-dimensional spacetime, and does not
include all the possible moduli of, for example, toroidal
compactification, since it assumes the torus is diagonal.) 

We will adopt this rolling radii Ansatz for the metric and the dilaton.
However, the point of this paper is to study solutions of type II
theories and M--theory by turning on non-trivial Ramond-Ramond
form-fields. Can we understand how these will modify the moduli space
description? Such solutions were first considered in the context of
eleven-dimensional supergravity by Freund and Rubin
\cite{freund_rubin}. Solutions with a non-trivial Neveu-Schwarz
two-form field, usually in four dimensions, have also been discussed
by various authors \cite{freund,antoniadis,tseytlin,be_fo,axion}.
Following Freund \cite{freund}, if we are to preserve
the structure of a spacetime split into a set of maximally symmetric
subspaces, the form fields must be chosen to span completely one or
more of the subspaces. They may or may not in addition have a
component in the time direction. Our central result will be to 
show that the effect of such a configuration is to introduce
new exponential potential terms into the
moduli space equations. In general the new equations describing the
evolution of the dilaton and the scale-factors can only be solved
exactly in certain special cases. We presented some examples of these
special cases in a previous letter \cite{letter}. (One example
appeared at almost the same time in paper by Kaloper \cite{kal}.) A subsequent
paper by L\"u {\it et al} \cite{lu}, gave a broad class of further 
exact solutions. Recently, cosmological solutions with Ramond forms
obtained from black hole solutions have also been studied~\cite{be_fo,rudi}.
The present paper provides a general framework for
analyzing cosmologies with Ramond forms, the Neveu-Schwarz form
and the cosmological constant of type II theories as well as those in
eleven dimensions. 

For all the cases we can solve exactly, our general result is that the
moduli-space potential which arises when non-trivial form-fields are
excited, significantly effects the structure of the solution. 
When all the subspaces are flat, the potential is operative over a
particular, finite part of the evolution; 
at the extremes, the solution becomes pure geodesic 
motion and we return to the simple Kaluza-Klein Kasner-type solutions with
some subspaces expanding, some contracting. (There is consequently
always either an initial or a final curvature singularity.) Thus the
effect of the form fields is to interpolate between two different
Kaluza-Klein solutions. As we have mentioned, the expansion or
contraction in the Kaluza-Klein solutions is controlled by a power
law. In general, for solutions with an initial singularity, one finds
that the rate of expansion is always sub-luminal and so there is no
inflation. On the other hand, those solutions with a final
singularity, just as in the pre-big-bang models \cite{pre_BB}, may exhibit
superinflation but are unphysical as stands because the inflation ends
in a curvature singularity. 

The solution can be very different, however, when we allow for curved
subspaces. The effect of the curvature is to introduce new terms into
the moduli space potential, with the particular new feature that, in
the case of a spherical subspace, the new term is always negative. This
can significantly change the singularity structure, in some cases giving
solutions where the curvature always remains finite. This suggest the
interesting possibility that there may exist inflating solutions which
are not forced to end in a singularity.

It is worth noting that there is actually a close connection
between cosmological solutions and those describing
$p$-branes, as was pointed out for the first time in
ref.~\cite{letter} and recently analyzed, starting from black hole
solutions, in~~\cite{rudi}.
In the single $p$-brane solution \cite{duff_rep}, all the fields
depend on the radial direction transverse to the brane. If the brane is
elementary the form field has a component in the radial direction. If
it is solitonic it spans a sphere around the brane, with no radial
component. Since our solutions depend on time alone, time must play
the role of the radial coordinate. Thus these solutions should correspond to
purely spacelike $p$-branes, spatial $(p+1)$-dimensional objects which
appear at one instant in time. Similarly, the two possible orientations
of the form field in our solution, with a time component or without,
correspond to the elementary and solitonic branes respectively. There
is really one subtlety in this interpretation, at least when all the
subspaces are flat. A standard $p$-brane
solution in $D$ dimensions has a $(D-p-1)$-dimensional transverse
space, composed of a radial direction and a $(D-p-2)$-sphere, whereas, 
in the cosmological solutions, time is the only transverse direction. However
it is well known that the number of effective transverse directions
can be reduced by stacking arrays of parallel $p$--branes together
\cite{branestack}. In the
limit of smooth distribution of branes the fields no longer depend on
the transverse coordinate along the direction of the array but only
on the coordinates transverse to both the brane and the array. Thus we
see that our cosmological solutions should really correspond to a
distribution of parallel, spacelike branes evenly spread throughout
space, though again appearing at only one instant in time. Solutions
with multiple form fields excited then correspond to distributions of
intersecting spacelike brane \cite{multibrane}. It is known that
$p$--branes can be considered as interpolating solutions between
different purely gravitational backgrounds. Therefore, the result that our
cosmological solutions interpolate between pure Kaluza-Klein
solutions supports the above analogy. The analogy breaks down however
with regard to supersymmetry. Unlike the case of conventional
$p$-brane solutions, we found no cosmological solutions which
preserved any supersymmetries. We will see that the actual form of the
solutions is closer to the black $p$-brane \cite{blackp,duff_rep} and
other non-extremal brane solutions \cite{lu_pope}. 

If one allows non-flat subspaces the $p$-brane analogy can become even
clearer. Behrndt and F\"orste \cite{be_fo} and, more recently, Poppe
and Schwager \cite{rudi} have obtained cosmological solutions directly
from black $p$-brane solutions by going to a regime where roles of the
time and radial coordinate are exchanged. Such solutions correspond in
our formalism to allowing a single spherical subspace. This subspace
together with the time direction then constitute the space transverse
to a single $p$-brane. This raises the possibility of finding new
cosmological solutions of this form which correspond directly to
single black $p$-brane states. 

The organization of the paper is as follows. In section two we give
the general structure of the solutions. Starting with the form of the
effective action, we give the metric Ansatz and discuss the possible
elementary- and solitonic-type Ans\"atze for the form fields. We then
describe the action in moduli space which reproduces the spacetime
equations of motion, and discuss its form. In section three, we give
the exact general solution corresponding to exciting a single form
field with flat subspaces, and discuss 
its structure as an interpolating solution between
two pure Kaluza-Klein solutions, together with an explicit example. In
section four, we give a class of exact solutions with multiple forms
excited. These are solvable because they correspond to Toda
models. Again, giving explicit examples in two distinct cases with
flat subspaces, we
discuss the interpolating form of these solutions. In section five, we
explain how to incorporate the case of curved spatial subspaces 
into our formalism. A class of solutions appears to be
related to black $p$-branes. We give, as an example, the solution of
Behrndt and F\"orste \cite{be_fo}, and discuss the possibility that those
solutions which are related to black $p$-branes do not begin or
end in a curvature singularity. We present some very brief concluding
remarks in section six. 


\section{General framework}

In this section we present a general framework for finding
cosmological solutions with non-trivial form fields. While here the
discussion will be as general as possible, we briefly discuss the same
analysis in terms of simple, illustrative example in section 3.2.
The starting point for our investigation is the following effective action 
\be
 \bar{S} = \int d^Dx\sqrt{\bar{g}}\left[e^{-2\f}(\bar{R}+4(\partial\f )^2
           -\frac{1}{12}H^2)-\sum_r \frac{1}{2(\d_r+1)!}F_r^2-\L \right]
 \label{string_action}
\ee
with the $D$--dimensional string frame metric $\bar{g}_{MN}$, the dilaton
$\f$, the NS 2--form $H$ and a number of RR $\d_r$--forms $F_r=dA_r$. We
also allow for a cosmological constant $\L$ which appears in the massive
extension of IIA supergravity~\cite{romans}. Assuming this origin, it
is restricted to be positive, $\L >0$. Note also that this cosmological
constant behaves much like a RR form in that is does not couple to the
dilaton in the string frame. In this respect
it is quite different from a ``normal'' cosmological constant appearing in an
effective action of noncritical dimension which, like the NS field, couples
to $\exp (-2\f )$.

The above action can account for a wide range of cosmological solutions
in type II theories (where we usually have $D=10$ in mind) and, if the
dilaton is set to zero, in $D=11$ supergravity. For simplicity we have
kept only the kinetic terms for the form fields. In general, in both
type II theories and eleven dimensional supergravity there are
additional terms involving the coupling of form fields. We shall
assume throughout this paper that for the configurations with which we
are concerned, these terms do not contribute to the equations of
motion. This will always be true if we only excite one orientation 
(in the sense given below) of a single form. It is not necessarily
true when multiple forms are excited. 

\vspace{0.4cm}

In order to give a physical description of our solutions, we prefer to work
in the canonical Einstein frame metric $g_{MN}$ which is related to
the string frame metric by a conformal rescaling $g_{MN}=\exp (-4/(D-2)\f
)\bar{g}_{MN}$. The corresponding action reads
\be
 S=\int d^Dx\;\sqrt{-g}\left[ R-\frac{4}{D-2}(\partial\f )^2-\sum_r
    \frac{1}{2(\d_r+1)!}e^{-a(\d_r)\f}F_r^2-\L e^{-a_\L \f}\right]
    \label{action}
\ee
where the NS field $H$ has been included in the sum over $F_r$. It is
distinguished from the other forms by the dilaton couplings $a(\d_r)$ given
by
\be
 a(\d_r) = \left\{ \ba{cll} \frac{8}{D-2}&{\rm NS}&{\rm 2-form}\\
                      \frac{4\d_r -2(D-2)}{D-2}&{\rm RR}&\d_r{\rm -form}
                \ea\right.\; . \label{p_rr}
\ee
The coupling $a_\L$ for the cosmological constant
\be
 a_\L = -\frac{2D}{D-2}  \label{p_lambda}
\ee
equals the negative of the one for a RR $(D-1)$--form. This reflects
the fact that it couples to the 8--brane solution~\cite{bergs_8brane}
of IIA supergravity. 

\vspace{0.4cm}

The type of solutions we consider are characterized by a space split
into $n$ flat subspaces, each of them characterized
by a scale factor $\a_i$. We concentrate on {\em flat} subspaces since this
covers already a wide range of solutions and seems to be appropriate for our
main purpose, namely to investigate the effect of the forms. The effect of
{\em curved} maximally symmetric subspaces can, however, be easily incorporated
into the formalism to be developed in this section. We will discuss this in
section six. In the flat case, the corresponding metric is given by
\be
 ds^2 = -N^2(\t )d\t^2+\sum_{i=0}^{n-1}e^{2\a_i(\t )}d{\bf x}_i^2
 \label{metric}
\ee
where $d{\bf x}_i^2$ is the measure of a $d_i$-dimensional flat maximally
symmetric subspace and $\sum_{i=0}^{n-1}d_i=D-1$. For solutions with
this structure, the
dilaton should depend on time only, $\f =\f (\t )$. This Kaluza--Klein-type
Ansatz is about the simplest allowing for the cosmologically key properties
of homogeneity and isotropy as well as for an ``external'' and ``internal''
space. 

For a ten dimensional theory, the simplest choice is to split up the
space into two subspaces ($n=2$) 
with $d_0=3$ and $d_1=6$. Then the $d_0=3$ part could be interpreted as
the spatial part of ``our'' 4--dimensional  space--time with an evolution
described by the scale factor $\a_0$. The other six dimensions would form an
internal space with a modulus $\a_1$. Clearly, one is free to split these
six dimensions even further and to consider, for example, a situation
with $n=3$ and $(d_0,d_1,d_2)=(3,4,2)$. From the cosmological point of
view one might even allow for a further split of the
3--dimensional space such as $(d_0,d_1,d_2)=(1,2,6)$ as long as
the $1+2$--dimensional space approaches a 3--dimensional isotropic space
at late times. 

As discussed in the introduction, this Kaluza-Klein-type Ansatz is
closely related to the form of $p$--brane solutions in supergravity
theories. Consider the simple case of two subspaces with $d_0=p+1$ and
$d_1=D-p-2$. For the standard $p$--brane solution, spacetime is split
into a radial direction, transverse to the brane, upon which all the
fields depend, a flat subspace describing the $(p+1)$--dimensional
worldvolume of the brane and hence including a timelike direction,
and a spherical subspace surrounding the brane. In our cosmological
solution the fields depend only on the timelike direction and so time
is the analog of the radial coordinate in the $p$-brane solutions. 
The $d_0$ subspace then becomes the analog of the brane worldvolume,
though it is now completely spacelike, while the $d_1$ subspace
becomes the sphere surrounding the brane. As described in the
introduction, since we usually take the $d_1$ space to be flat rather
than spherical, probably a more exact analogy is with a distribution
of spacelike $p$-branes, rather than a single brane. More complicated
cases, where the space is split into a number of subspaces, are
analogous to multiple intersecting $p$-brane solutions
\cite{multibrane}. 

The choice of subspaces is important in fixing the possible
structure of the antisymmetric tensors fields. 
The symmetry of the above metric allows for two different Ans\"atze
for the form fields which we call ``elementary'' and ``solitonic'' in
analogy to the two types of $p$--brane solutions. They are
characterized by the following nonvanishing components of the field
strength. 
\begin{itemize}
 \item elementary~: if $\sum_i d_i=\d_r$ for some of the spatial subspaces $i$
                    we may set
 \be
  (F_r)_{0\m_1...\m_{\d_r}} = A_r(\a )\, f_r'(\t )\,\e_{\m_1...\m_{\d_r}}\; ,
  \quad A_r(\a ) = e^{ -2\sum_i d_i\a_i } \label{elementary}
 \ee
 where $\m_1...\m_{\d_r}$ refer to the coordinates of these subspaces,
 $f_r(\t )$ is an arbitrary function to be fixed by the form field
 equation of motion, and the prime denotes the derivative with
 respect to $\t$. With raised indices, the symbol
 $\e^{\m_1...\m_{\d_r}}$ takes the values 0 or 1 and is completely 
 antisymmetric on all $\d_r$ indices. Note 
 that the sum over $i$ in the exponent runs only over
 those subspaces which are spanned by the form.
\end{itemize}
An elementary form can therefore extend over one or more of the
subspaces only if its degree matches the total dimension of these spaces.
Consider for example the RR three-form of type IIA supergravity. If the space
is split up as $(d_0,d_1)=(3,6)$ it fits into the 3--dimensional subspace
and the above general Ansatz specializes to
$F_{0\m_1\m_2\m_3}=\exp (-6\a_0)f'(\t )\e_{\m_1\m_2\m_3}$ where
$\m_1,\m_2,\m_3$ refer to the coordinates of this subspace. Taking
$(d_0,d_1,d_2)=(1,2,6)$ allows for a similar Ansatz but the form
now extends over the $1+2$--dimensional space and the exponential in the
Ansatz changes to $\exp (-2\a_0-4\a_1)$.
\begin{itemize} 
 \item solitonic~: if $\sum_i d_i=\d_r+1$ for some of the spatial subspaces $i$
                   we may allow for
 \be
  (F_r)_{\m_1...\m_{\d_r+1}} = B_r(\a )\; w_r\; \e_{\m_1...\m_{\d_r+1}}\;
 ,\quad
  B_r(\a ) = e^{ -2\sum_i d_i\a_i } \label{solitonic}
 \ee
 where $\m_1...\m_{\d_r+1}$ refer to the coordinates of these subspaces and
 $w_r$ is an arbitrary constant. As for the elementary Ansatz, the sum
 over $i$ in the exponent runs over the subspaces spanned by the
 form. It is easy to check that this Ansatz, already solves the form
 equation of motion. 
\end{itemize}
Note that in contrast to the elementary Ansatz, the solitonic field strength
does not have a time index. Therefore the matching condition for the
dimensions differs. Given, for example, a split $(d_0,d_1)=(3,6)$ one has to
use a  2 form instead of a three-form to fit into the 3-dimensional subspace.
The above Ansatz then reads
$F_{\m_1\m_2\m_3}=\exp (-6\a_0)w\e_{\m_1\m_2\m_3}$.

Why did we use the terms ``elementary'' and ``solitonic'' for the above
Ans\"atze? This is again motivated by a close analogy to $p$--brane
solutions. Consider again the simple split into two subspaces,
$d_0=p+1$ and $d_1=D-p-2$, where $d_0$ is the analog of the $p$-brane
worldvolume, and $d_1$ is the analog of the sphere surrounding the
brane. There are two types of the simple brane solutions. Each is a 
source of a form field, but the field is aligned differently in the
two cases. For the ``electrically charged'' elementary brane, the
form field is aligned in the hyperplane of the brane together with the
radial direction. This has an analog in our elementary Ansatz with
the form aligned in the $d_0$ subspace together with the time
direction. For the ``magnetically charged'' solitonic brane, the form
field is aligned in the sphere surrounding the brane, which in turn
has an analog in our solitonic Ansatz with the form aligned simply in
the $d_1$ subspace. 

\vspace{0.4cm}

Having specified the form of our Ansatz, we now look to solve the
equations of motion derived from the
action~\refs{action}. However, it is in fact easy to show that,
under a very mild restriction on the structure of forms to be discussed
below, the resulting equations of motion can be obtained from a 
reduced Lagrangian which depends only on $\a_i$, $\f$, $N$ and
$f_r$, each as functions of $\t$. The Lagrangian is given by
\be
 {\cal L} = E\left[\sum_{i=0}^{n-1}d_i\a_i'^2-
            \sum_{i,j=0}^{n-1}d_id_j\a_i'\a_j'
            +\frac{4}{D-2}\f '^2+V_e-N^2V_s\right]
 \label{lagrangian}
\ee
with
\bea
 V_e &=& \frac{1}{2}\sum_{r}A_r(\a )e^{-a(\d_r)\f}{f_r'}^2 \nn\\
 V_s &=& \frac{1}{2}\sum_{r}B_r(\a )w_r^2e^{-a(\d_r)\f}+\L
         e^{-a_\L\f} \label{Ves} \\
 E &=& \frac{1}{N}e^{\sum_{i=0}^{n-1}d_i\a_i }\; . \nn
\eea
In the definitions of the potentials $V_e$ and $V_s$, the sum over $r$ is
understood to run over all the elementary and solitonic configurations
which have been chosen according to the given rules. In general, this even 
includes the possibility of having a superposition of two of these
configurations for one form. However, in this case the nonvanishing
components of the two configurations should differ in at least
two spatial directions in order to obtain the correct equations of motion
{}from Lagrangian~\refs{lagrangian} (otherwise the equations of motion
contain an additional cross term which cannot be derived from
eq.~\refs{lagrangian}). We will adopt this fairly mild restriction on
the structure of the forms thereby gaining the advantage of having
a simple description of our models in terms of a reduced Lagrangian. 

The equations of motion for the functions $f_r$ to be derived from
eq.~\refs{lagrangian} read
\be
 \frac{d}{d\t}\left( EA_re^{-a(\d_r)\f}f_r'\right) = 0 \; .
\ee
The first integrals are
\be
 f_r' = v_rE^{-1}A_r^{-1}e^{a(\d_r)\f} \label{fr}
\ee 
where $v_r$ are integration constants. Equation~\refs{fr} can be used to
eliminate $f_r'$ from the elementary potential $V_e$. This then
reduces the problem to solving the remaining equations of motion now
given purely in terms of $\a_i$, $\f$ and $N$. 

In fact we find that the remaining equations can also be derived from
a simple reduced Lagrangian. This Lagrangian describes the motion of a
particle in a ``moduli space'' the coordinates of which are the scale
factors together with the dilaton. We find $E$ is then the metric for
the particle worldline in the moduli space. To see this structure, we
first introduce the notation $\bal=(\a_I)=(\a_i,\f )$ for a general
point in the moduli space. We also define a particular metric on the
moduli space $G_{IJ}$ by 
\bea
 G_{ij}&=&2(d_i\d_{ij}-d_id_j)\nn \\
 G_{in}&=&G_{ni}=0 \label{G}\\
 G_{nn}&=&\frac{8}{D-2}\; .\nn 
\eea
The equations of motion for $\bal$ and $N$ following from
eq.~\refs{lagrangian} then take the simple form
\bea
 \frac{d}{d\t}\left( EG\bal '\right)+E^{-1}\frac{\partial U}{\partial\bal}
  &=&0 \label{al_eom} \\ 
 \frac{1}{2}E{\bal '}^TG\bal '+E^{-1}U &=& 0\; . \label{N_eom}
\eea
The quantity $U$ is given by
\be
 U=e^{2\sum_{i=0}^{n-1}d_i\a_i}\left( \frac{1}{N^2}V_e+V_s\right)
 \label{pot_def}
\ee
where $f_r'$ in $V_e$ has been replaced using eq.~\refs{fr}. Clearly,
these equations of motion can be derived from the simple Lagrangian
(it is convenient to make the change of variables from $N$ to $E$),
\be
 {\cal L} = \frac{1}{2}E{\bal '}^TG\bal '-E^{-1}U \label{dan_lag}
\ee
We note from its definition \refs{G}, that the
metric on the moduli space has Minkowskian signature $(-++\cdots +)$,
thus we are really free to interpret \refs{dan_lag} as the worldline
Lagrangian for particle moving in the moduli space. The first term is
kinetic, while $U$ defines a potential in the moduli space. Further, 
$E$ is the metric on the particle worldline. It encodes the gauge
freedom that we have not yet chosen a particular time parameterization
to describe the evolution of the spacetime. As usual its equation of
motion \refs{N_eom} is a constraint.

It is useful to rewrite the potential $U$ in a more systematic
way as
\be
 U = \frac{1}{2}\sum_{r=1}^{m}u_r^2\exp (\bq_r \cdot\bal ) \label{U}
\ee
where the sum runs over all elementary and solitonic configurations as well
as a possible cosmological constant term. The constants $u_r$ represent
the integration constants $v_r$ in eq.~\refs{fr} for elementary forms,
the constants $w_r$ in the Ansatz~\refs{solitonic} for solitonic
forms or a cosmological constant. The type of each term is specified
by the vectors $\bq_r$ which can be read off from eqs.~\refs{elementary},
\refs{solitonic}, \refs{Ves} and the definition~\refs{pot_def}. For an
{\em elementary} $\d$--form they are given by
\be
 \bq^{\rm (el)} = (2\e_id_i,a(\d ))\; ,\quad \e_i=0,1\; ,
 \quad \d =\sum_{i=0}^{n-1}\e_id_i \label{q_el}
\ee
with $\e_i=1$ if the form is nonvanishing in the subspace $i$ and
$\e_i=0$ otherwise. For type II theories the dilaton couplings $a(\d )$ are
given in eq.~\refs{p_rr} and~\refs{p_lambda}. To account for the $D=11$
case (or constant dilaton solutions) we may just set $a(\d )=0$.

Let us give an example at this point. An elementary IIA RR 3 form, put
into the first subspace of a $D=10$ space split with $(d_0,d_1)=(3,6)$, is
characterized by a vector $\bq = (6,0,-1/2)$. It generates a
potential term in~\refs{U} which depends on $\a_0$ and the dilaton but not
on $\a_1$. More generally, a potential term describing
the effect of an elementary form depends only on those scale factors
which correspond to subspaces spanned by the form. Since the
entries $q_i$ of $\bq$ are always positive, the potential tends to
drive the scale factors for these subspaces to smaller values. Therefore
these subspaces tend to be contracting or at least less generically
expanding than others.

The situation for a {\em solitonic} $\d$--form is in some sense complementary.
It is specified by a vector
\be 
 \bq^{\rm (sol)} = (2\tilde{\e}_id_i,-a(\d ))\; ,\quad
  \tilde{\e}_i\equiv 1-\e_i
 =0,1\; ,\quad \tilde{\d}\equiv D-2-\d =\sum_{i=0}^{n-1}\tilde{\e}_id_i
 \label{q_sol}
\ee
with $\tilde{\e}_i=1$ if the form vanishes in the subspace $i$ and
$\tilde{\e}_i=0$ otherwise. For example, a solitonic IIB RR 2 form in the
first subspace of a space split as $(d_0,d_1)=(3,6)$ is specified
by $\bq = (0,12,1)$. It generates a potential term in~\refs{U} which depends
on $\a_1$ and the dilaton but not on $\a_0$. More generally, in contrast to
the elementary case, the potential term now depends on those scale factors
corresponding to subspaces {\em not} spanned by the form.

Finally, a cosmological constant is characterized by
\be
 \bq^{(\L )} = \left( 2d_i,\frac{2D}{D-2}\right)\label{q_lambda}\; .
\ee
Note that for all these vectors $\sum_{i=0}^{n-1}q_i>0$, a fact which
we will use later on. The moduli space metric allows us to define a
natural scalar product on the space of vectors $\bq$.. It is the
product $\bq\cdot\bal$ that appears in the potential $U$. Thus $\bq$
is naturally a covariant rather than a contravariant vector in the moduli
space, so we define 
\be
 <\bq_1 ,\bq_2 > = \bq_1^TG^{-1}\bq_2 \label{s_prod}
\ee
with the inverse of $G$ given by
\bea
 (G^{-1})_{ij} &=& -\frac{1}{2(D-2)}+\frac{1}{2d_i}\d_{ij} \nn \\
 (G^{-1})_{in} &=& (G^{-1})_{ni} = 0 \label{Gin}\\
 (G^{-1})_{nn} &=& \frac{D-2}{8} \; . \nn
\eea
Since the metric $G$ has Minkowskian signature, we can 
distinguish between space- and time-like vectors $\bq$. As we will see,
the structure of the solutions depends crucially on this distinction.

\vspace{0.4cm}

Generically, the models specified by the eqs.~\refs{al_eom}, \refs{N_eom}
and~\refs{U} cannot be solved. A complete solution, however, can be found
if the potential $U$ consists of one exponential term only or if contact
with Toda theory can be made. These two cases will be discussed in section
3 and 4. An interesting observation at this stage is that the potential
$U$ may contain exponentials with different signs in front of the dilaton
$\f$ and therefore may have dilaton minima at finite values. It is an
important question how this influences the dynamics of the dilaton,
in particular since the stabilization of the dilaton is one of the
major problems of string cosmology. This will be analyzed more
completely elsewhere~\cite{wip}.


\section{Solutions with one potential term} 

In this section, we will analyze models with just one form turned
on (or a non-zero cosmological constant). The form may be elementary or
fundamental and there may be any number of subspaces. All of these
cases correspond to a potential 
\be
 U = \frac{1}{2}u^2\exp (\bq \cdot\bal ) \label{U1}
\ee
where $u^2>0$. Here $\bq$ is one of the vectors specified at the end of the
last section. We will start by giving the general form of the
solution and then give a simple example in section 3.2.

One way of solving the equations of motion for a potential~\refs{U1} is
to use the gauge freedom in the definition of the time coordinate. We
can always choose a gauge such that 
\be
 N=\exp ((\bd -\bq )\cdot\bal ) \label{gauge}
\ee
where we introduce a vector giving the subspace dimensions$\bd =
(d_i,0)$. This implies $E = \exp (\bq \cdot\bal )$ and the following
set of equations for $\bal$ 
\bea
 \frac{d}{d\t}(GE\bal ')+\frac{1}{2}u^2\bq &=& 0 \nn \\
 \frac{E}{2}{\bal^T}'G\bal ' +\frac{1}{2}u^2 &=& 0\; \label{eom_one}.
\eea
In this form they can be integrated immediately, leading to the general
solution
\be
 \bal = \bc\ln |\t_1 -\t |+\bw\ln\left(\frac{s\t}{\t_1-\t}\right)+\bk
 \label{sol1}
\ee
where
\be
 \bc = \frac{2G^{-1}\bq}{<\bq ,\bq >}\; . \label{c}
\ee
The sign $s=\pm 1$ is determined by
$s = {\rm sign}(<\bq ,\bq >)$~\footnote{Here we disregarded the somewhat
marginal possibility that $\bq$ is a null vector, i.~e. $<\bq ,\bq >=0$.}
and $\bw$, $\bk$ are integration constants subject to the constraints
\bea
 \bq \cdot\bw &=& 1 \nn \\
 \bw^TG\bw &=& 0 \label{cons1}\\
 \bq \cdot \bk &=& \ln\left(\frac{u^2|<\bq ,\bq >|}{4}\right)\; . \nn
\eea
$\t_1$ is a free parameter which we can take to be positive. The opposite
choice $\t_1<0$ leads to a shift in the range of $\t$ which does not affect
the physics of the solution. The range of $\t$ should be chosen to ensure
a positive argument of the second logarithm in eq.~\refs{sol1}. This depends
on the sign of $<\bq ,\bq >$ and we have the two cases 
\be
 \ba{ccc}
  0<\t <\t_1&{\rm for}&<\bq ,\bq >\; > 0\\
  \t <0\;{\rm or}\; \t>\t_1&{\rm for}&<\bq ,\bq >\; < 0
 \ea\; . \label{ranges}
\ee
Which of these cases is actually realized in type II models? Using
the vectors $\bq$ given in the end of section 2, we find for
a solitonic or elementary $\d$ form (or a cosmological constant which
is similar to a RR $(D-1)$--form)
\be
 <\bq ,\bq > = \frac{D-2}{8}a(\d )^2+\frac{2}{D-2}\d\tilde{\d}=\left\{
               \ba{cll} 4&&{\rm NS} \\
                        \frac{D-2}{2}&&{\rm R}
               \ea \right. \label{qq}
\ee
which is always positive. Also the $D=11$ 3 form leads to
a positive result, as can be seen from the above formula by setting $a(\d )=0$.
We conclude that, in the present context, we are dealing with spacelike
vectors $\bq$ only, and we have $0<\t<\t_1$.

\vspace{0.4cm}

For concreteness, let us give an example of how to compute the
coefficients $\bc$ for a given model. Consider an elementary $\d$
form in $D=10$ and a spacetime split into two subspaces with dimensions
$d_0=\d$ and $d_1=D-\d-1=\tilde{\d}+1$. Then from eq.~\refs{G} we find
the metric  
\be
 G=\left(\ba{ccc} -2\d (\d -1)&-2\d (\tilde{\d}+1)&0\\
                  -2\d (\tilde{\d}+1)&-2\tilde{\d}(\tilde{\d}+1)&0\\
                   0&0&1\ea\right) \label{G_ex}
\ee
and from eq.~\refs{Gin} its inverse
\be
 G^{-1}= \left(\ba{ccc} \frac{1}{2\d}-\frac{1}{16}&-\frac{1}{16}&0\\
                        -\frac{1}{16}&\frac{1}{2(\tilde{\d}+1)}-
                        \frac{1}{16}&0\\
                        0&0&1\ea\right)\; .
 \label{Gin_ex}
\ee
Inserting this into eq.~\refs{c}, together with the vector
$\bq = (2\d ,0,a(\d ))$ characterizing the elementary $\d$ form, we find
the following coefficient $\bc$ of the first term in the
solution~\refs{sol1}
\be
 \bc = \left(\frac{\tilde{\d}}{2(\d+\tilde{\d})},
             -\frac{\d}{2(\d+\tilde{\d})},\frac{a(\d )}{2}\right)\; .
 \label{c_ex}
\ee

\vspace{0.4cm}

The above example provides useful insight into the relation of our solutions
to $p$--brane solutions. As discussed in the last section, we would
expect this example to correspond to an elementary $(\d -1)$-brane
solution. Eq.~\refs{sol1} has been written in a form which makes this
relation transparent. The first term in eq.~\refs{sol1} is the analog
of an extremal, BPS, p-brane solution. To see this, we remind the
reader that a brane BPS solutions is characterized by a
proportionality of all $\a_I$ to a certain harmonic function $h$; that
is $\a_I\sim h$. The $(\d -1)$-brane constants of
proportionality~\cite{duff_rep} are exactly reproduced by
eq.~\refs{c_ex}. The second term in eq.~\refs{sol1} represents the
analog of nonextremal extensions which also have been studied in the
$p$-brane case~\cite{blackp,duff_rep,lu_pope}. The
difference here is that the second term cannot be set to zero since
$w_I=0$ for all $I$ is incompatible with the
constraints~\refs{cons1}. Another possibility is to choose $\t_1 =0$
which by eq.~\refs{ranges} works for timelike vectors $\bq$
only. Unfortunately, as we saw, such vectors do not occur within
$D=10$ type II and $D=11$ supergravity. By analogy with $p$--brane
solutions one would therefore conclude that the solution~\refs{sol1}
does not preserve any of the supersymmetries of these theories. This
can, in fact, be checked explicitly for certain examples by using the
supersymmetry transformations of the gravitino.

\vspace{0.4cm}

So far, our solutions have been expressed in terms of the time parameter
$\t$ which is defined by the gauge choice~\refs{gauge} for $N(\t )$.
For a discussion of the cosmological properties of our models,
however, we should reexpress them in terms of the comoving time $t$,
that is in the gauge where the $N=1$. This can
be done by integrating the defining relation $dt=N(\t )d\t$. 
The explicit expression for $N(\t )$ is given 
by inserting the solution~\refs{sol1} into the gauge fixing
equation~\refs{gauge} for $N(\t )$, which gives 
\be
 N = \exp ((\bd -\bq )\cdot\bk )|\t_1-\t |^{-x+\D -1}||\t |^{x-1} \label{N}\\
\ee
with
\bea
 x &=& \bd \cdot\bw \label{x}\\
 \D &=& \bd \cdot\bc = 2\frac{<\bd ,\bq >}{<\bq ,\bq >} \; . \label{Delta}
\eea
Note that the quantity $x$ depends on the specific choice of
integration constants $\bw$ whereas $\D$ is a fixed number for a given model
(once the dimensions $\bd$ of the subspaces and the Ansatz for
the form encoded in $\bq$ have been chosen). Depending on the values of
$x$ and $\D$, the gauge parameter $N$ may have singularities at $\t =0$ and
$\t =\t_1$. This determines the allowed range in the comoving time $t$ as we
will discuss in detail in the next subsection.

Another quantity which is of importance in discussing the physical
content of our solutions is the scalar curvature $R$. For the
solution~\refs{sol1} it is given by
\be
R \sim |\t_1 -\t |^{2(x-\D )}|\t |^{-2x}P_2(\bw ,\bal ,\t ) \label{R}
\ee
where $P_2$ is a second order polynomial in $\t$ which we will not need
explicitly. We have also omitted an unimportant constant of proportionality
which depends on the integration constants. The first two factors in this
equation indicate potential singularities at $\t = 0$ and $\t =\t_1$,
depending on $x$ and $\D$ as in the case of the
gauge parameter $N$. However, in contrast to singularities in $N$, such
singularities are true coordinate independent curvature singularities.
They will be further discussed in the next subsection. 


\subsection{Cosmology of solutions with spacelike $\bq$--vectors}

As already mentioned, the case of spacelike $\bq$-vectors 
is the most important in our context since
all vectors $\bq$ appearing within the $D=10$ type II theories and $D=11$
supergravity are spacelike. Even under this restriction on $\bq$, our
solution~\refs{sol1} covers a large number of different models.
In this section, we will discuss the cosmological properties which can be
extracted from these models in general. A concrete illustrating example
will be given in the next subsection.

Recall that the singularity structure 
of the solution~\refs{sol1} is determined by the
quantities $x$ and $\D$ defined in eq.~\refs{x} and~\refs{Delta}. What values
are actually allowed for these quantities?
{}From $<\bd ,\bq >=-\sum_{j=0}^{n-1}q_j/2(D-2)$ and $\sum_{j=0}^{n-1}q_j>0$
it follows from eq.~\refs{Delta} that $\D <0$. Though we are not able to
give a general proof, the parameter $x$, which unlike $\D$ depends on
the parameters of the solution, turns out to be either $x<\D$
or $x>0$ in all specific examples we considered. This divides the set
of initial conditions into two disconnected subsets corresponding to
two classes of solutions with different properties. 

We begin our discussion of these properties by analyzing the allowed ranges
in comoving time $t$. Recall from eq.~\refs{ranges} that the time parameter
$\t$ is always in the range $0<\t <\t_1$ since we have $<\bq ,\bq >\;
>0$. The singularity structure of the gauge parameter $N$ in
eq.~\refs{N} then shows that this range is mapped to the following
ranges in $t$ 
\be
 \t\rightarrow t\in\left\{\ba{clll}
       \left[ -\infty ,t_1\right]&{\rm for}\; x<\D<0\; ,&(-)\;{\rm branch} \\
       \left[ t_0,+\infty\right]&{\rm for}\; x>0>\D\; ,&(+)\;{\rm branch}
       \ea\right.\; .
\ee
Here $t_0$ and $t_1$ are two finite unrelated values that appear as
integration constants from integrating $dt =N(\t )d\t$. Thus we have found
two disconnected branches corresponding to asymptotically positive and
negative time ranges.

Let us next discuss the scalar curvature in each branch. We start with
the $(-)$ branch. As inspection of eq.~\refs{R} shows, the curvature
vanishes as $t\rightarrow -\infty$ ($\t\simeq 0$) since $x<\D <0$. With
increasing time $R$ grows and, finally, the system runs into the curvature
singularity at $t=t_1$ ($\t =\t_1$) since the power $2(x-\D )$ of the
first term in eq.~\refs{R} is negative. Therefore, classically the
$(-)$ solution cannot be continued beyond this point.

In the $(+)$ branch the situation is similar but reversed in time. At
$t=t_0$ ($\t =0$) we find a singularity since $x>0$ in this branch.
The solution cannot be extended into the past. As $t\rightarrow\infty$
($\t\simeq\t_1$) the curvature behaves smoothly and approaches zero.

Though generically correct, the above argument has a loophole.
For very specific values of the initial parameters $\bw$, the polynomial
$P_2$ is proportional to $\t$ or $\t_1-\t$ so that it can cancel against
one of the first two factors in eq.~\refs{R} which cause the singularity.
If $|x|$ is sufficiently small, the singularity may disappear completely.
For the $(-)$ branch this is realized if $w_n=c_n$ and $x\geq -1/2$. For
the $(+)$ branch it occurs if $w_n=0$ and $x\leq 1/2$.
This phenomenon is quite similar to what happens in the curvature
singularity free WZW model of ref.~\cite{kou_lust}

\vspace{0.4cm}

So far, we have  considered quantities which provide information
about the behaviour of the total $D$ dimensional space only. Let us now turn
to the individual subspaces of dimension $d_i$. To analyze their
behaviour, we should calculate their respective Hubble parameters $H_i$ in
terms of the comoving time. In fact, it is possible to explicitly
express the comoving time $t(\t )$ in terms of hypergeometric functions.
It is, however, more instructive to look at the asymptotic regions $\t\simeq 0$
(corresponding to $t\rightarrow -\infty$ for the $(-)$ branch and
$t\simeq t_0$ for the $(+)$ branch) and $\t\simeq\t_1$
(corresponding to $t\simeq t_1$ for the $(-)$ branch and
$t\rightarrow\infty$ for the $(+)$ branch). In these regions the Hubble
parameters can be written as~\footnote{The dot denotes the derivative with
respect to the comoving time $t$.}
\be
 \bH \equiv \dot{\bal} = \frac{\bp}{t-t_s} \label{hubble}
\ee
with the constant expansion coefficients $\bp$ satisfying
\be
 \bp G\bp = 0\; ,\quad \bd \cdot\bp = 1\; .\label{kk}
\ee
The time shift $t_s$ depends on the asymptotic region and branch under
consideration.
The sign of $t-t_s$, however, is always well defined~: It is negative in
the $(-)$ branch and positive in the $(+)$ branch. If we combine the
two equations~\refs{kk} and use the explicit form of the metric
$G$ in~\refs{G} we find
\be
 \frac{4}{D-2}p_\f^2+\sum_{i=0}^{n-1}d_ip_i^2 = 1\; .\label{p_bound}
\ee
The explicit expressions for $\bp$ in terms of the integration constants
are 
\be
 \bp = \left\{\ba{cll} \frac{\bw}{x}&{\rm at}&\t\simeq 0 \\
                       \frac{\bw -\bc}{x-\D}&{\rm at}&\t\simeq\t_1
       \ea\right. \; . \label{p_expr}
\ee
They have been calculated using the general solution~\refs{sol1} and
the asymptotic limits of $N(\t )$ to be read off from eq.~\refs{N}.
The behavior of the Hubble parameters~\refs{hubble} along with eq.~\refs{kk}
indicates that the solutions behave asymptotically like those of pure
Kaluza--Klein theory with a dilaton. This can be seen by a comparison with
the solutions of ref.~\cite{mueller}. Therefore, one expects that the
potential $U$ provided by the form is effectively turned off in these
limits. In fact, inserting the general solution~\refs{sol1} into the
potential~\refs{U1}, we find
$U\sim (\t_1-\t )\t$ which implies that $U$ is effectively zero near
$\t\simeq 0$ and $\t\simeq \t_1$. The effect of the form is therefore
to generate a mapping $\bp (\t\simeq 0)\rightarrow\bp (\t\simeq\t_1)$
between two pure Kaluza--Klein states.

What do the above results mean for the evolution of the subspaces? We
consider the $(+)$ branch first. Remember that $t-t_s>0$ in this branch
so that from eq.~\refs{hubble} a positive $p_i$ results in expansion
and a negative $p_i$ in contraction. Moreover, the equation $\bd\cdot\bp = 1$
shows that at least one of the $p_i$ has to be positive. Consequently, at
least one of the subspaces has to be expanding. From eq.~\refs{p_bound} we
conclude that $|p_i|<1$ always. The expansion is therefore subluminal,
i.~e.~the scale factor $e^{\a_i}$ grows less quickly than the horizon
size $H_i^{-1}$. This behaviour is similar to a radiation or matter
dominated universe corresponding to $p_i=1/2$ and $p_i=2/3$, respectively.

The situation is completely different in the $(-)$ branch. Since
$t-t_s<0$, a positive $p_i$ results in contraction and a negative $p_i$
in expansion. Now we conclude from $\bd\cdot\bp =1$ that at least one
subspace must be contracting. Since we are in the negative time range,
eq.~\refs{hubble} shows that expansion ($H_i>0$) goes along with an
increasing $H_i$, that is, a shrinking horizon size. Such a behaviour
is also called superinflation since scales are stretched across the
horizon even more rapidly than in ``ordinary'' inflation where the horizon
size is approximately constant. It is very similar to the superinflating
expansion in pre--big--bang models~\cite{pre_BB}. However, it should be
stressed that it was obtained in the Einstein frame as opposed to the
string frame used in those models.

Our solutions allow various patterns of expanding and contracting spatial
subspaces. The details of the evolution depend on the partition $\{ d_i\}$,
the form and the subspaces it occupies and the initial conditions.
Examples with $3$ expanding and $6$ contracting spatial dimensions
as $t\rightarrow\infty$ exist, as were given in
ref.~\cite{letter}. Though this is by no means preferred in
the present context, its existence is not entirely trivial since some
models fail to reproduce such a ``desired'' situation. The effect of
the form can be quite dramatic. For example, it can reverse expansion
and contraction of two subspaces during the early asymptotic period into
its converse during the late period.

A ``preferred'' cosmological scenario suggested by these solutions consists
of a combination of the $(-)$ and the $(+)$ branch to account for inflation
as well as for a postinflationary subluminal expansion. It is remarkable
that such a pre--big--bang scenario can be realized in the Einstein frame.
The apparent shortcoming of this scenario is that the $(-)$ and the
$(+)$ branch constitute two different {\it a priori} unrelated
solutions. As for string frame pre--big--bang models, one might
argue~\cite{pre_BB} that scale factor (T) duality between the branches
provides the correct transition and some hints for this have been
found in a two dimensional toy model~\cite{rey}.


\subsection{An example}

Up to this point our discussion has been rather general. Let us now
illustrate the steps in our solution by giving a simple
example. It is not meant to represent the most realistic case in the
cosmological sense, but rather to explain our general results as
concretely as possible. 

We consider the following situation~: 10-dimensional spacetime is
split into two subspaces with $\bd = (d_0,d_1,0)=(3,6,0)$ and an
elementary IIA RR 3 form occupies the 3--dimensional subspace. This
implies the Ansatz 
\bea
 ds^2 &=& -N^2(\t )d\t^2 +e^{2\a_0}d\bx_0^2+e^{2\a_1}d\bx_1^2\nn \\
 F_{0\m_1\m_2\m_3} &=& e^{-6\a_0}f'(\t )\e_{\m_1\m_2\m_3} \\
 \f &=& \f (\t ) \nn
\eea
in accordance with the eqs.~\refs{metric}, \refs{elementary}.
The equations of motion for this example can be derived from the
Lagrangian
\be
 {\cal L} = E\left[-6{\a_0'}^2-30{\a_1'}^2-36\a_0'\a_1'+\frac{1}{2}
            {\f '}^2+V_e\right] \label{lag_ex}
\ee
with the elementary potential $V_e$ and $E$ given by
\be
 V_e = \frac{1}{2}e^{-6\a_0+\frac{1}{2}\f}{f'}^2\; ,\quad \label{V_ex}
 E = \frac{1}{N}e^{3\a_0+6\a_1} \; .
\ee
which come from the general equations~\refs{lagrangian} and
\refs{Ves}. Note that this Lagrangian depends on the function $f'$
which appears in the Ansatz for the form, but not on $f$
itself. Therefore the equation of motion for $f$ can be integrated to
give the first integral 
\be
 f' = uE^{-1}e^{6\a_0+\frac{1}{2}\f} \label{f_ex}
\ee
with an integration constant $u$. From the Lagrangian~\refs{lag_ex} we
can compute the equations of motion for $\a_0$, $\a_1$ and $\f$. Using
eq.~\refs{f_ex} to replace $f'$ in these equations, we arrive at
\bea
 \frac{d}{d\t}\left( E(-12\a_0'-36\a_1')\right) +3u^2E^{-1}e^{6\a_0-\f /2}
  &=& 0 \nn \\
 \frac{d}{d\t}\left( E(-36\a_0'-60\a_1')\right) &=& 0 \label{eom_ex} \\
 \frac{d}{d\t}\left( E\f '\right)-\frac{1}{4}u^2E^{-1}e^{6\a_0-\f /2}
   &=& 0\nn \; .
\eea
Let us compare these equations with the general ones given in the
moduli space formalism in \refs{al_eom},
\refs{N_eom} and \refs{U}. We see that they can be indeed written in this
form if we set
\be
 G = \left(\ba{ccc} -12&-36&0\\
                      -36&-60&0\\
                       0&0&1\ea\right)\; . \label{G_ex2}
\ee
and
\be
 U=\frac{1}{2}u^2e^{6\a_0-\frac{1}{2}\f} \; . \label{U_ex}
\ee
The matrix $G$ above is consistent with the general formula~\refs{G}
with $d_0=3$, $d_1=6$ and $D=10$. In eq.~\refs{U} we 
introduced a systematic way of writing the effective potential by
introducing a characteristic vector $\bq_r$ for each form. From
eq.~\refs{U_ex} we read off that this vector is given by
$\bq = (6,0,-1/2)$ for our example. This coincides with what one gets by
applying the general rule~\refs{elementary} to the breakup
$(d_0,d_1)=(3,6)$ and a $\d =3$ form in the 3--dimensional subspace. The
dilaton coupling $a(\d )$ for a RR 3--form needed in eq.~\refs{elementary}
follows from eq.~\refs{p_rr} to be $a(\d )=-1/2$.

In section 2 we also defined a scalar product~\refs{s_prod} on the space
of vectors $\bq$ using the inverse of $G$. From eq.~\refs{G_ex2} $G^{-1}$
is given by
\be
 G^{-1} = \left(\ba{ccc} \frac{5}{48}&-\frac{1}{16}&0\\
                         -\frac{1}{16}&\frac{1}{48}&0\\
                         0&0&1\ea\right) \; . \label{Gin_concrete}
\ee
which agrees with the general formula~\refs{Gin} for $(d_0,d_1)=(3,6)$ and
$D=10$. One can easily verify that $<\bq ,\bq > =\bq G^{-1}\bq = 4$.
Therefore $\bq$ is indeed a spacelike vector,
in agreement with the general result~\refs{qq} which showed that this is true
for all vectors obtained from type II forms.

\vspace{0.4cm}

The main problem in solving the system of equations~\refs{eom_ex} is the
existence of two different exponentials, one coming from $E$,
eq.~\refs{V_ex}, the other coming from the effective potential $U$,
eq.~\refs{U_ex}. Fortunately, we have a gauge freedom (time
reparameterization invariance) encoded in $N$ which we can use to get
rid of one of the exponentials. In the next section we will use this
freedom to set $E=1$. Here, however, we choose the another possibility,
namely to gauge away the potential by setting $E=\exp (6\a_0-\f /2)$.
Given the definition of $E$ in eq.~\refs{V_ex}, this implies
\be
 N=\exp (-3\a_0+6\a_2+\f /2) \label{N_ex}
\ee
in accordance with the general formula~\refs{gauge} for $\bd =(3,6,0)$ and
$\bq = (6,0,-1/2)$. With this choice, the equations of motion~\refs{eom_ex}
turn into
\bea
 \frac{d}{d\t}\left( e^{6\a_1-\f /2}(2\a_1'+6\a_2')\right) &=& \frac{u^2}{2}
  \nn \\
 \frac{d}{d\t}\left( e^{6\a_1-\f /2}(3\a_1'+5\a_2')\right) &=& 0 \\ 
 \frac{d}{d\t}\left( e^{6\a_1-\f /2}\f '\right) &=& \frac{u^2}{2} \nn \\
 e^{6\a_1-\f /2}(6{\a_1'}^2+36\a_1'\a_2'+30{\a_2'}^2-{\f '}^2) &=&
 \frac{u^2}{2} \nn \; .
\eea
This is consistent with the general form~\refs{sol1} found for models with one
term in the potential. Taking appropriate linear combinations of the
first three equations we can derive an equation for the remaining
exponent $6\a_0-\f /2$, which can be solved. Then $\a_0,\a_1,\f$ can be
expressed in terms of this solution. In this way one arrives at
a general solution of the form~\refs{sol1} with coefficients $\bc$
given by
\be
 \bc = \left(\frac{5}{16},-\frac{3}{16},-\frac{1}{4}\right)\; .\label{c_ex2}
\ee
and the following constraints on the integration constants
\bea
 6w_1-\frac{1}{2}w_3 &=& 1 \nn \\
 12w_1^2+72w_1w_2+60w_2^2 &=& w_3^2 \label{cons1_ex} \\
 6k_1-\frac{1}{2}k_3 &=& \ln (u^2) \; .\nn
\eea 
One may arrive at the same result by inserting $\bq = (6,0,-1/2)$ and
$G^{-1}$ from eq.~\refs{Gin_concrete} into the general formulae~\refs{c},
\refs{cons1}. Recall that the time parameter $\t$ is restricted
by $0<\t <\t_1$.

To discuss the cosmology of these solutions we must perform a
transformation to comoving time $t$. To do this, we need the explicit form of
the gauge parameter $N$ which we find by inserting the solution~\refs{sol1}
with~\refs{c_ex2}, \refs{cons1_ex} into eq.~\refs{N_ex}
\be
 N=e^{-3k_0+6k_1+k_2/2}|\t_1-\t |^{-x+\D -1}|\t |^{x-1} \label{Nsol_ex}\; .
\ee
Here $x=3w_0+6w_1$ and $\D =-3/16$. The quantities $x$, $\D$ have been
generally defined in eq.~\refs{x}, \refs{Delta} and their values can be easily
reproduced by inserting $\bd = (3,6,0)$ and $\bc$ from eq.~\refs{c_ex2}.
The range of comoving time obtained by integrating $dt=N(\t )d\t$ over
$0<\t <\t_1$ crucially depends on the singularities in $N$.
Eq.~\refs{Nsol_ex} shows that there are potential singularities at $\t =0$
and $\t =\t_1$. Their appearance is controlled by the value of $x$.

Let us therefore analyze which values of $x$ are allowed. The first two
constraints~\refs{cons1_ex} may be solved to express, for example, $w_0$
and $w_1$ as a function of $w_2$. This shows that $x$ depends on one
free parameter only. Furthermore, since the second constraint~\refs{cons1_ex}
is quadratic in $w_I$, we find two branches satisfying $x<\D = -3/16$ and
$x>0$, respectively. We refer to these two branches as the $(-)$ and
the $(+)$ branch. From eq.~\refs{Nsol_ex} we see that $0<\t <\t_1$ is
indeed mapped to the comoving time ranges given in eq.~\refs{ranges};
that is to $\left[ -\infty ,t_1\right]$ for the $(-)$ branch and to
$\left[ t_0,\infty\right]$ for the $(+)$ branch ($t_0,t_1$ are integration
constants). Moreover, the scalar curvature~\refs{R} has a future timelike
singularity in the $(-)$ branch and a past timelike singularity in the
$(-)$ branch. Both types of solutions are therefore not extendible.

\vspace{0.4cm}

Information about the evolution of the two subspaces and the dilaton
can be obtained form the respective Hubble parameters $\bH =\dot{\bal}$
written as a function of comoving time. They can be calculated if
$\t\simeq 0$ or $\t\simeq\t_1$ since $N$ in eq.~\refs{Nsol_ex} can be
integrated in these limits. Doing this for our example by using
eq.~\refs{sol1}, \refs{c_ex2}, \refs{cons1_ex} and $N$, $\t (t)$
calculated from eq.~\refs{Nsol_ex}, we find $\bH$ to be of the
Kaluza--Klein form~\refs{hubble}, \refs{kk}. The expansion coefficients
$\bp$ depend on the integration constants $\bw$ as in eq.~\refs{p_expr}.

In fact, using the first two constraints~\refs{cons1_ex} we may rewrite
$\bp$ as a function of $w_2$ only, as we did for the parameter $x$ before.
The asymptotic expansion properties of our example therefore depend on
one free parameter only. Instead of giving the explicit formulae,
which are not particularly enlightening, let us give a graphical 
representation of $\bp =\bp (w_2)$. We concentrate on the $(+)$ branch (the
expansion coefficients in the $(-)$ branch can be worked out analogously)
where $t_0<t<\infty$ and the asymptotic regions are characterized by
$t\simeq t_0$ ($\t\simeq 0$) and $t\rightarrow\infty$ ($\t\simeq\t_1$). 
The results are given in fig.~1 ($t\simeq t_0$) and fig.~2
($t\rightarrow\infty$).
In both figures $|p_0|,|p_1|<1$ always, which illustrates our general
result that the expansion in the $(+)$ branch is always subluminal
(cf.~eq.~\refs{p_bound}).\\[-0.5cm]
\centerline{\epsfbox{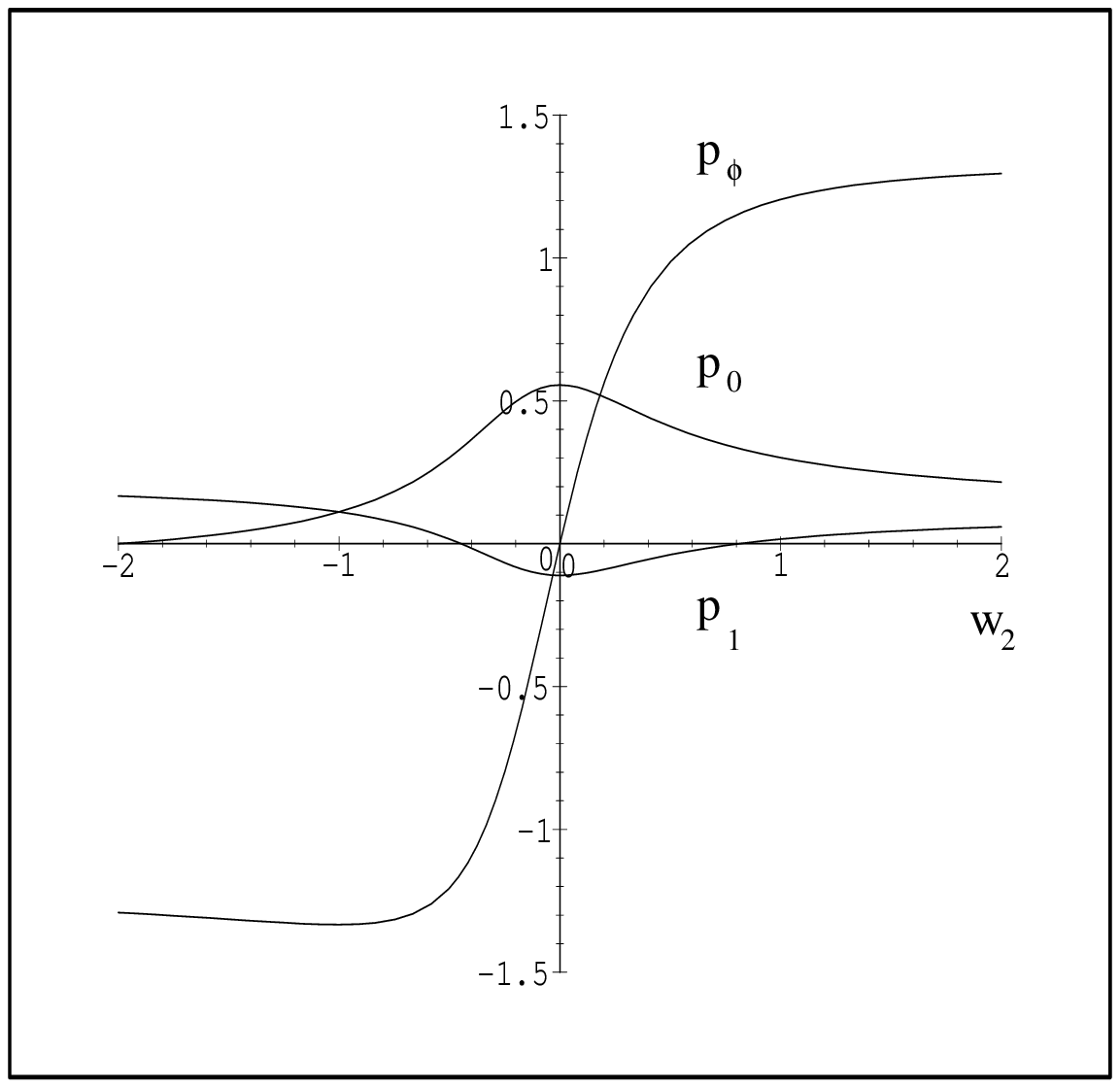}}
\centerline{\em Fig 1: Expansion coefficients for the
            ($+$) branch at $t\simeq t_0$.}
\vskip 0.4cm
We see that an early expansion of the 3--dimensional subspace
($p_0>0$ in fig.~1) is turned into a contraction as $t\rightarrow\infty$
($p_0<0$ in fig.~2) for a wide range in $w_2$. This can be understood from
the $\a_0$ dependence of the effective potential~\refs{U_ex}. Moreover, the
6--dimensional space is always expanding as $t\rightarrow\infty$
($p_1>0$ in fig.~2). In a more realistic model, such an expansion should be
stopped by, for instance, a nonperturbative potential for the modulus $\a_1$.
\centerline{\epsfbox{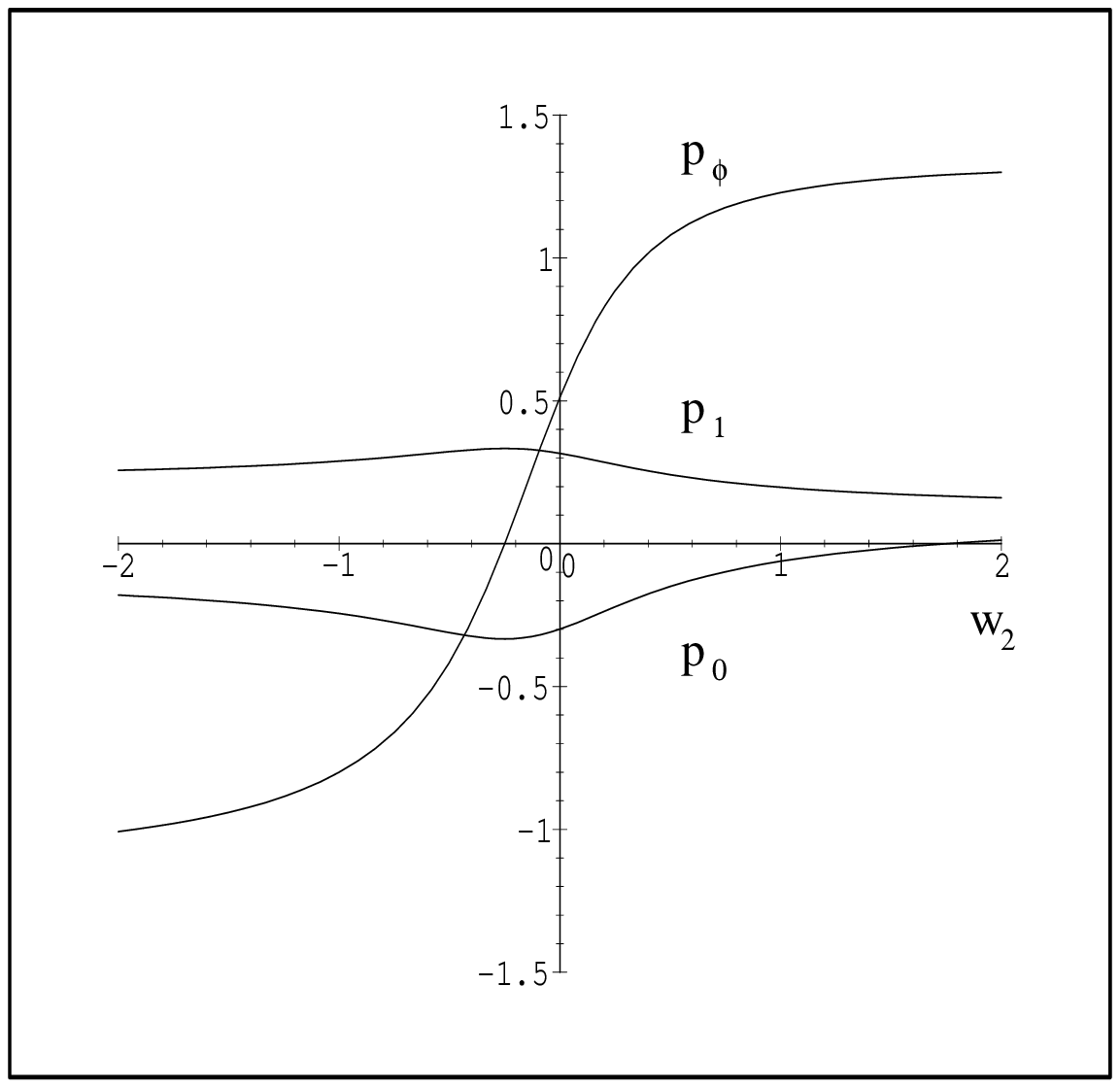}}
\centerline{\em Fig 2: Expansion coefficients for the ($+$) branch at
            $t\rightarrow\infty$.}
\vskip 0.4cm
%


\section{More general potentials -- Toda models}

So far, we have explicitly solved the case with just one form turned on.
As already mentioned, models with two or more forms are only soluble if
they correspond to Toda models. In this section we will establish the
relation to Toda models, and discuss two classes of solutions with
some examples. 

One recalls that the main result of section 2 was that the
problem of finding cosmological solutions could be reduced to solving
for the motion of a particle in a moduli space with a particular
potential. The coordinates in the moduli space were an $n+1$
component vector $\bal = (\a_i ,\f )$ containing the scale factors
$\a_i$, $i=0,...,n-1$ of the $n$ maximally symmetric subspaces in
which we divided the $D$--dimensional space and the dilaton.
The dynamics of $\bal$ were then described by the particle 
equations of motion~\refs{al_eom}, \refs{N_eom} with the potential
given by~\refs{U}. The inverse metric on the moduli space~\refs{Gin}
defined a natural scalar product $<\ ,\ >$~\refs{s_prod} for covariant
vectors living in the moduli space (such as $\bq$). We also recall
that the metric had Minkowskian signature. To make contact with Toda
models we would like to write all the moduli-space equations of motion
in terms of the scalar product. To this end we introduce a
``lowered-index'' covariant vector corresponding to the contravariant
coordinate vector $\bal$ by 
\be
 \bbe = G\bal.
\ee
Lets us also choose to work in the gauge
\be
 N = \exp (<\bd ,\bbe >) \; .\label{gauge_gen}
\ee
which implies that $E=1$. In this gauge, the moduli space
Lagrangian~\refs{dan_lag} written in terms of $\bbe$ reduces to a form
${\cal L}_0$. The constraint~\refs{N_eom} can then be expressed as the
vanishing of the corresponding Hamiltonian ${\cal H}_0$ so that the
whole system is described by 
\bea
 {\cal L}_0 &=& \frac{1}{2}<\bbe ',\bbe '>-U \nn \\
 {\cal H}_0 &=& \frac{1}{2}<\bbe ',\bbe '>+U = 0 \label{lag}\\
 U &=& \frac{1}{2}\sum_{r=1}^{m} u_r^2\exp (<\bq_r,\bbe >)\; .
\eea
The prime denotes the derivative with respect to the time parameter $\t$.
Recall that each of the $m$ terms in the potential $U$ corresponds to
a certain Ansatz for a form specified by the respective vector $\bq_r$,
$r=1,...,m$.

\vspace{0.4cm}

A general solution for this system can be found if it
corresponds to a Toda model~\cite{kostant}; that is, if the matrix
\be
 A_{rs} = \frac{<\bq_r,\bq_s>}{q} \label{Cartan}
\ee
can be identified with the Cartan matrix of a semi--simple Lie group $G$ for
some constant $q$~\footnote{If the group is semisimple $q$ can be different
for each simple factor.}. Note that $A$ is an $m\times m$ matrix, where $m$
is the number of forms excited, each specified by a different $\bq_r$. 

Let us briefly show that this property does not depend on the
particular coordinate system in moduli space we have chosen.
Under a linear coordinate transformation
$\bal\rightarrow\bar{\bal}=P\bal$, 
the covariant vectors $\bbe$ and $\bq_r$ transform as 
$\bbe\rightarrow\bar{\bbe}={P^{-1}}^T\bbe$ and
$\bq_r\rightarrow\bar{\bq}_r={P^{-1}}^T\bq_r$, while the new inverse
metric is given by $\bar{G}^{-1}=PG^{-1}P^T$. Clearly all scalar
products are unchanged. Thus, the matrix
$A_{rs}$ is invariant under such an operation and the property of
being (or not being) a Toda model is coordinate independent. In particular,
the transition from the Einstein to the string frame discussed in the
beginning of section 2 can be described by such a change of coordinates in
moduli space. This shows that our decision to work in the Einstein frame
does not affect the essentials of the present discussion. A similar remark
applies to duality transformations within the class of models considered
here, which we also expect to act as linear transformations on $\bbe$.

\vspace{0.4cm}

Are there really such Toda models among the examples one obtains from
type II theories? To answer this question systematically, and to give a
simple criterion to decide in each specific case, we will now compute
the possible entries of the matrix $A$. We consider two forms of
degree $\d_1$ and $\d_2$ with vectors $\bq_1$, $\bq_2$ as
defined in eq.~\refs{elementary} and \refs{solitonic}.
{}From eq.~\refs{s_prod}, their scalar product is given by
\bea
 <\bq_1^{(\rm el,NS)},\bq_2^{(\rm el,NS)}> &=& 2\d_{12} \nn \\
 <\bq_1^{(\rm el,R)},\bq_2^{(\rm el,R)}> &=& 2\d_{12}-\d_1-\d_2
  +\frac{D-2}{2} \label{el_el} \\
 <\bq_1^{(\rm el,R)},\bq_2^{(\rm el,NS)}> &=& 2\d_{12}-2 \nn
\eea
and
\bea
 <\bq_1^{(\rm sol)},\bq_2^{(\rm sol)}> &=&
   <\bq_1^{(\rm el)},\bq_2^{(\rm el)}>-2 \label{sol_sol}\\
 <\bq_1^{(\rm el)},\bq_2^{(\rm sol)}> &=&
    - <\bq_1^{(\rm el)},\bq_2^{(\rm el)}> \label{el_sol}
\eea
where $\d_{12}=\sum_{i=0}^{n-1}\e_i^{(1)}\e_i^{(2)}d_i$ is the
``spatial overlap'' of the two forms, that is the dimension of the
subspace in which both forms are non-zero. As before,
the cosmological constant of massive IIA is analogous to a RR $(D-1)$--form.
The above expressions show that the entries of the matrix $A$ are small
integer numbers (at least for $D=10$) of either sign which are just the
right properties to construct a Cartan matrix. For a number of models $A$
can indeed be identified with a Cartan matrix. Explicit examples of
this have been discussed in ref.~\cite{letter}. Here we are mainly interested
in the general structure of the solutions.


\subsection{Orthogonal vectors $\bq$}

A simple class of soluble models~\cite{russian_guys} is characterized by
orthogonal vectors $\bq_r$; that is, $<\bq_r,\bq_s>=0$ for $r\neq s$.
Then the matrix $A$ is diagonal and can, via appropriate rescalings, be
identified with the Cartan matrix of $SU(2)^m$.

Before we solve the equations of motion, let us show that such a situation
can indeed occur within type II theories. We consider two forms of type IIA,
namely an elementary 3--form specified by $\bq_1$ and a solitonic 1--form
specified by $\bq_2$ occupying {\em different} subspaces. In the
language of eq.~\refs{el_el} that means $\d_1=3$, $\d_2=1$ and
$\d_{12}=0$ (no overlap). Then from eq.~\refs{el_sol} and \refs{el_el} we have
$<\bq_1,\bq_1>=<\bq_2,\bq_2>=4$ and $<\bq_1,\bq_2>=0$ and, hence, we
have orthogonal vectors $\bq_r$.

\vspace{0.4cm}

Let us now solve the equations of motion. It is useful to introduce an
orthonormal basis $\boe_i$, $i=0,...,n$ in moduli space satisfying
\be
 <\boe_i ,\boe_j >= \eta_i\d_{ij} \label{basis}
\ee
with $\eta_0=-1$ and $\eta_i=1$ for $i>0$. In the following, we will use
indices $i,j,...=0,...,n$ to label these basis vectors. Remember that the
scalar product has a Minkowskian signature so that one of these vectors,
$\boe_0$, is normalized to $-1$. We have assumed that the vectors $\bq_r$,
$r=1,...,m$ are orthogonal. This allows us to identify some of the basis
vectors $\boe_i$, with a normalized version of the $\bq_r$
\be
 \boe_r = \frac{-\bq_r}{q_r}\; ,\quad q_r =\sqrt{|<\bq_r ,\bq_r>|}
          \; ,\quad r=1,...,m\; . \label{q_basis}
\ee
Note here, that the $\bq_r$ are orthogonal {\em spacelike} vectors. This
implies that their number (the number of forms) is less than the dimension
of the moduli space; that is, $m\leq n$. Furthermore, none of the $\bq_r$
can be identified with the timelike basis vector $\boe_0$.
The basis vectors~\refs{basis} therefore fall into two groups, those which
can be identified with forms and those which cannot. Correspondingly, we
partition the index $i=0,...,n$ as $i=(a,r)$, where we use
the same index $r=1,...,m$ as for the forms to label the basis
vectors proportional to $\bq_r$ and the index $a=0,m+1,...,n$ to
label the other basis vectors.

We expand the ``dynamical'' vector $\bbe$ in terms of the basis
\be
 \bbe = \sum_{i=0}^{n}\r_i\boe_i\; . \label{expansion}
\ee
with (time dependent) expansion coefficients $\r_i(\t )$. 
Inserting this expansion into ${\cal L}_0$ given in eq.~\refs{lag}, we
find a complete decoupling in the eigenmodes $\r_i$ and the
resulting equations of motion 
\bea
 \r_r''&=&\frac{1}{2}q_ru_r^2e^{-q_r\r_r} \\
 \r_a''&=& 0 \; .
\eea
Their solution given by
\be
 \ba{lll}
  \r_r &=&  q_r^{-1} \ln (g_r) \\
  \r_a &=&  k_a(\t -\t_a)
 \ea \label{rho_sol}
\ee
with where we define
\be
 g_r = K_r\cosh^2 (z_r)\; ,\quad K_r=\frac{u_r^2}{k_r^2}\; ,\quad
 z_r = \frac{1}{2}|k_r|q_r(\t -\t_r)\; , \label{gr}
\ee
Here $\t$ runs over the full real axis and $k_i$ and $\t_i$ are integration
constants. Bringing everything together, the solution for the original
vector $\bal$ reads 
\be
 \bal = \sum_{i=0}^{n}\r_iG^{-1}\boe_i \label{sol2}
\ee
with $\r_i$ as indicated above. 

The application to an explicit example
is straightforward. First one should calculate the matrix $G^{-1}$ from
eq.~\refs{Gin}. Then one determines the vectors $\bq_r$ specifying the forms
{}from eq.~\refs{elementary}, \refs{solitonic} and calculates the
basis vectors $\boe_r$, $r=1,...,m$ via eq.~\refs{q_basis}. Finally,
one extends this subset of basis vectors to a full orthogonal basis
of the moduli space satisfying eq.~\refs{basis}. All these quantities
inserted into eq.~\refs{sol2} provide the explicit solution.  

\vspace{0.4cm}

So far we have not considered the Hamiltonian constraint
${\cal H}_0=0$ in eq.~\refs{lag}. For the above solution it leads to
\be
 {\cal H}_0 = \frac{1}{2}\sum_{i=0}^{n}\eta_ik_i^2 = 0\; . \label{cons2}
\ee
Correspondingly, the constants $k_i$ can be interpreted as the
``energy'' contribution of the modes $\r_i$.
To understand the meaning of the ``time shifts'' $\t_r$ it is instructive
to compute the potential $U$ for the above solution
\be
 U = \frac{1}{2}\sum_{r=1}^{m}u_r^2g_r^{-1}=\frac{1}{2}\sum_{r=1}^{m}
     k_r^2\left[\cosh(z_r)\right]^{-2}\; . \label{U2}
\ee
It indicates that the form $r$ is operative around $\t\simeq\t_r$ and
switched off otherwise so that we expect asymptotic Kaluza--Klein
behavior at $\t\rightarrow\pm\infty$. Suppose we time order the
time shifts $\t_r$; that is, $\t_{r_1}\leq\t_{r_2}\leq
...\leq\t_{r_m}$. Then the system
starting out at $\t\rightarrow -\infty$ will go through a sequence of
``transformations'' each induced by one of the forms before it reaches the
asymptotic region at $\t\rightarrow\infty$. If the time differences
$\t_{r_{i+1}}-\t_{r_i}$ are all of order one, then the forms will act almost
simultaneously. However, if there is a long interval between two
of the time shifts, for example $\t_{r_2}-\t_{r_1}\gg 1$, the potential is
effectively turned off during much of the intermediate period
$\t_{r_1}<\t<t_{r_2}$. Depending on the number of long intervals, we may
have up to $m-1$ of these intermediate Kaluza--Klein regions. 

Though our solutions cannot be generally 
expressed in terms of the comoving time in closed form, this can be 
done in the Kaluza-Klein regions. For the gauge parameter $N$ we find from
eq.~\refs{gauge_gen} and \refs{expansion}
\be
 N=\exp\left(\sum_i\D_i\r_i\right)\; ,\quad \D_i=<\bd ,\boe_i>\; .
 \label{Delta_r}
\ee
Using the explicit expression for $\r_i$, eq.~\refs{rho_sol}, this can be
approximated in any one of the Kaluza--Klein regions as
\be
 N\simeq N_0e^{\n\t}\; ,\quad \n=\sum_{a}k_a\D_a+\sum_{r}(\pm )
                               |k_r|\D_r\; . \label{N2}
\ee
Note that there is a sign freedom in the second term of the second equation.
This freedom serves to specify the asymptotic region for which $N$ is
computed. Let us concentrate on the asymptotic region specified by
$\t_{r_i}<\t<\t_{r_{i+1}}$. Then all terms with $r=r_1,...,r_i$ get
a $+$ sign and all terms with $r=r_{i+1},...,r_m$ get a $-$ sign.
More intuitively, a term $r$
gets a $+$ sign if the $r$th form has been ``passed'' and a $-$ sign otherwise.
Of particular importance are the two parameters $\n_\pm$ corresponding to
$\t\rightarrow\pm\infty$ which determine the relevant range of the comoving
time $t$. We get $\n_+$ from the above formula for $\n$ by taking all signs
to be $+$ and $\n_-$ by taking all signs to be $-$. From eq.~\refs{N2},
the quantities $\n_\pm$ do not seem to be restricted in range. We have,
however, to consider the Hamiltonian constraint~\refs{cons2} on the
integration constants $k_i$. Using this constraint, we find for all
explicit examples that $\n_+$ and $\n_-$ always have the same sign. Then
integration of $dt=N(\t )d\t$ with $N$ as in eq.~\refs{N2} leads to two
unconnected branches with
\be
 \t\rightarrow t\in\left\{\ba{cll}
 \left[ -\infty ,t_1\right] \; ,&\n_\pm <0&(-)\;{\rm branch} \\
 \left[ t_0,+\infty\right] \; ,&\n_\pm >0&(+)\;{\rm branch} \ea \right.\; .
 \label{branch2}
\ee
The Hubble parameters in each Kaluza--Klein region take the standard
form \refs{hubble}, with the expansion coefficients $\bp$ expressed in terms
of the integration constants as
\be
 \bp = \frac{1}{\n}\left[\sum_{a}k_aG^{-1}\boe_a+
       \sum_{r}(\pm )|k_r|G^{-1}\boe_r\right]\; . \label{p2}
\ee
The signs in the second term have to be chosen according to the same rule
as above. Depending on the choice of these signs, eq.~\refs{p2} specifies
the expansion parameters in each of the up to $m+1$ asymptotic regions.
To get this result we have used the definition of $\bH$ in
eq.~\refs{hubble}, the solution~\refs{sol2} and $N$, $t(\t )$ to be
calculated from eq.~\refs{N2}.
Clearly, these coefficients fulfill the two equations~\refs{kk}
characteristic for a Kaluza--Klein region. Therefore, the discussion
of cosmological properties given in section 3 based on these equations
applies here as well. In particular, the expansion in the $(-)$ branch is
superinflating for $p_i<0$ and the expansion in the $(+)$ branch is always
subluminal since $|p_i|<1$.

\vspace{0.4cm}

In summary, we have found the direct generalization of the simple case
with one form~: the solutions split into two different branches $(+)$, $(-)$
with properties similar to those discussed in section 3. We also have
two Kaluza-Klein regions, one at early and one at late time, in 
each respective branch.
If all time shifts $\t_r$ are of the same order, these two regions are the
only Kaluza-Klein regions, as in 
the case of one form. The transition, however, is then
generated by the combined effect of all forms. If there are long
intervals between the $\t_r$ we may have up to $m-1$ intermediate
Kaluza--Klein regions, where $m$ is the number of forms. In each
region the forms are effectively turned off. A transition is
generated when one or more of the forms become operative.

\subsection{A Toda example}

Let us illustrate the general results above by an explicit example.
We break up the space as $\bd = (3,2,4,0)$ and consider an elementary
IIA 3 form in the $d_0=3$ subspace and a solitonic IIA 1 form in
the $d_1=2$ subspace. From the eqs.~\refs{elementary} and
\refs{solitonic} we find the corresponding vectors
$\bq_1 = (6,0,0,-1/2)$ for the 3 form and $\bq_2 = (6,0,8,3/2)$ for
the 1 form. In the notation of eq.~\refs{el_el} we therefore have
$\d_1=3$, $\d_2=1$ and $\d_{12}=0$. Inserting these quantities into the
eqs.~\refs{el_el}, \refs{sol_sol} and \refs{el_sol} we find
$<\bq_1,\bq_1>=<\bq_2,\bq_2>=4$ and $<\bq_1,\bq_2>=0$. This shows that
our example indeed corresponds to an orthogonal model.
By using the inverse matrix
\be
 G^{-1} = \left(\ba{cccc}\frac{5}{48}&-\frac{1}{16}&-\frac{1}{16}&0\\
                         -\frac{}{16}&\frac{3}{16}&-\frac{1}{16}&0\\
                         -\frac{1}{16}&-\frac{1}{16}&\frac{1}{16}&0\\
                         0&0&0&1
          \ea\right)
\ee
obtained from eq.~\refs{Gin} this can also be verified explicitly.
Apart from $G^{-1}$ we need the set $\boe_i$, $i=0,1,2,3$ of orthogonal
vectors in
moduli space to get the explicit solution from eq.~\refs{sol2}.
The two basis vectors $\boe_1$, $\boe_2$ in the direction of the forms follow
easily from eq.~\refs{q_basis} by normalizing $\bq_1$ and $\bq_2$.
The two remaining vectors $\boe_0$, $\boe_3$ should be chosen to complete
this to a basis of the moduli space. A possible choice for the complete
basis is
\bea
 \boe_0 &=& (6,2,8,0) \nn \\
 \boe_1 &=& \left( -3,0,0,\frac{1}{4}\right) \nn \\
 \boe_2 &=& \left( -3,0,-4,-\frac{3}{4}\right) \\
 \boe_3 &=& \sqrt{6}\left( 1,0,2,-\frac{1}{4}\right) \nn \; .
\eea
It is straightforward to verify that this set of vectors indeed fulfills the
normalization condition~\refs{basis}. Inserting $G^{-1}$ and $\boe_i$ into
eq.~\refs{sol2}, along with the functions $\r_i$, gives the explicit
solution for our example. It follows from eq.~\refs{rho_sol} that
the functions $\r_i$, $i=0,1,2,3$ read
\be
 \r_r = \frac{\ln (g_r)}{q_r}
\ee
for $r=1,2$ where $g_r$ is given in eq.~\refs{gr}, and
\be
 \r_a = k_a(\t -\t_a)
\ee
for $a=0,3$. The effective potential $U$, eq.~\refs{U2}, specializes to 
\be
 U = \frac{1}{2}\left( k_1^2\cosh^{-2}(z_1)+k_2^2\cosh^{-2}(z_2)\right)
 \label{U3}
\ee
for our example, where $z_r$ has been defined in eq.~\refs{gr}. Note
that the first term in this potential is related to the 3 form and the
second one to the 1 form.

For a discussion of cosmological properties the numbers
$\D_i$ defined in eq.~\refs{Delta_r} are needed (see for example
eq.~\refs{N2}). From the above basis vectors they are easily computed
to be $\D_0 = -1$, $\D_1 = 3/16$, $\D_2=7/16$ and $\D_3 =-3\sqrt{6}/16$.
Now we have explicitly given all quantities required to compute the
gauge parameter $N$, eq.~\refs{N2}, and the expansion parameters $\bp$,
eq.~\refs{p2}, as a function of the integration constants $k_i$. Note
that the Hamiltonian constraint~\refs{cons2} should be used to eliminate 
one of these integration constants.

\vspace{0.4cm}

Let us discuss the various cases that can occur within our model.
Recall first that each form $r=1,2$ is operative at $\t\simeq\t_r$ only,
where $\t_r$ are free integration constants. The dashed line in fig.~3 shows
the potential~\refs{U3} for $\t_1=-1$, $\t_2=1$, so that the
difference in the time shifts is 
of order one. It can be seen that the effect of the two forms
cannot be separated in time in this case. We have only two asymptotic
regions, $\t\rightarrow -\infty$ and $\t\rightarrow \infty$. As an example,
we want to compute the quantities $\n_\pm$, which specify the behaviour
of the gauge parameter $N$, in these regions. Using the constants $\D_i$
given above we find from eq.~\refs{N2}
\be
 \n_- = -k_0-\frac{3\sqrt{6}}{16}k_3-\frac{3}{16}|k_1|-\frac{7}{16}|k_2|
\ee
for the asymptotic region $\t\rightarrow -\infty$. Correspondingly, for
the region $\t\rightarrow \infty$ we find
\be
 \n_+ = -k_0-\frac{3\sqrt{6}}{16}k_3+\frac{3}{16}|k_1|+\frac{7}{16}|k_2|\; .
\ee
Note that the $r=1,2$ terms are both negative for $n_-$
and both positive for $n_+$.
This is in agreement with the general description of the sign choice, which
we have given below eq.~\refs{N2}.
The Hamiltonian constraint~\refs{cons2} reads
\be
 {\cal H}_0 = -k_0^2+k_1^2+k_2^2+k_3^2 = 0\label{cons5}
\ee
and can be used to eliminate, for example, $k_0$ from the above expressions
for $\n_\pm$. Doing this, it can be seen explicitly, that $\n_+$ and
$\n_-$ always have the same sign, independent on the values of $k_1,k_2,k_3$.
As stated before, this leads to the two branches indicated in
eq.~\refs{branch2}. The computation of the expansion coefficients $\bp$ in
eq.~\refs{p2} can be carried out in a similar way. We will, however, not
give the explicit formulae here.

If $|\t_2-\t_1|\gg 1$ another intermediate Kaluza--Klein region appears.
This is illustrated by the solid line in fig.~3 where we have plotted the
potential $U$, eq.~\refs{U2}, for $\t_1=-10$ and $\t_2=10$. For this choice
of $\t_r$ the system in the intermediate region has already been influenced
by the 3 form but not yet by
the 1 form (of course one can have the converse situation by taking
$\t_1\gg\t_2$). Correspondingly, for the expansion coefficients in
eq.~\refs{p2} one should take the $+$ sign for the $r=1$ term and the
$-$ sign for the $r=2$ term. The expression for $\n$ in this intermediate
region therefore reads
\be
 \n = -k_0-\frac{3\sqrt{6}}{16}k_3+\frac{3}{16}|k_1|-\frac{7}{16}|k_2|\; .
\ee

\vspace{0.4cm}

So far, we have carried out the discussion in terms of the time parameter
$\t$. Upon mapping to the comoving time $t$ via eq.~\refs{branch2}, we
have two different branches, the $(+)$ branch and the $(-)$ branch.
The above discussion therefore applies to both branches. In particular, in
both branches we may have either two or three Kaluza--Klein regions,
depending on the values of the time shifts $\t_1$ and $\t_2$. Though
we have not computed the expansion coefficients $\bp$ explicitly, we know
{}from their general properties~\refs{kk} that the evolution in each of
these Kaluza--Klein regions will be similar to what we have described
in section 3 for one form only. \\
\centerline{\epsfbox{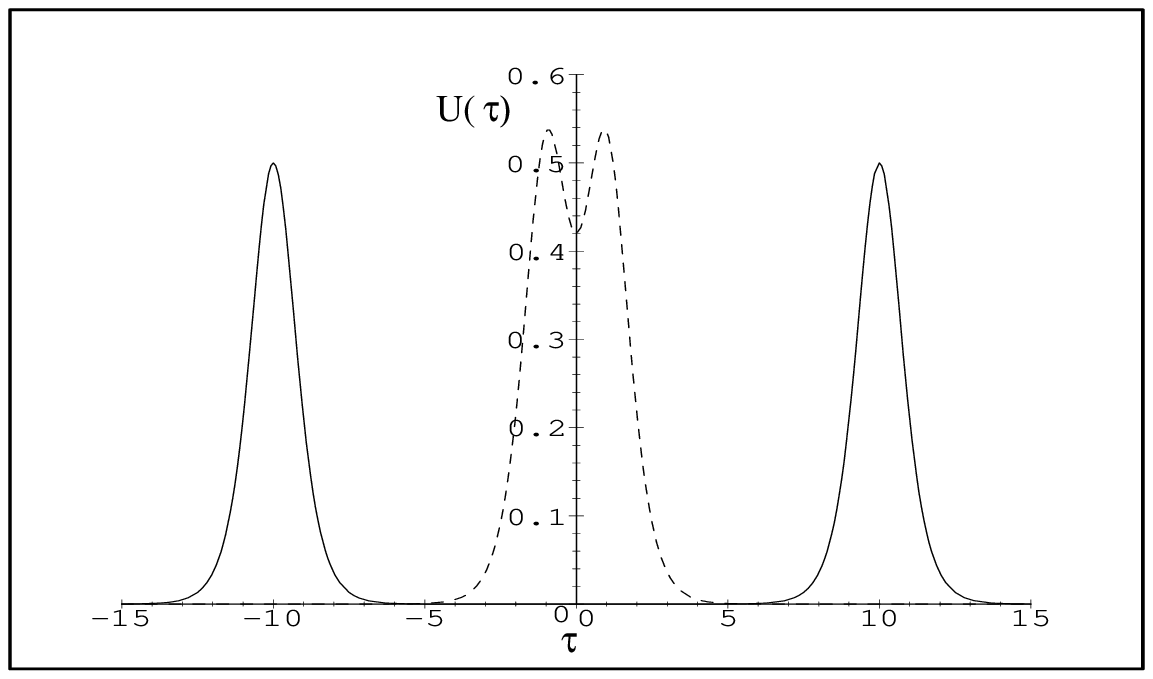}}
{\em Fig 3: Effective potential $U$ as a function of the time 
parameter $\t$ for the example described in the text. The solid
line corresponds to $\t_1=-10$, $\t_2=10$, the dashed line to
$\t_1=-1$, $\t_2=1$. For simplicity we have taken $k_1=k_2=1$.}
\vskip 0.4cm


\subsection{Nonorthogonal vectors $\bq$}

We now discuss the general case where the $\bq_i$ vectors are not
orthogonal. As in the previous subsection we start with an
expansion~\refs{expansion} for $\bbe$ in terms of a basis $\boe_i$,
$i=0,...,n$ in moduli space. Again, we would like to identify some of
the basis vectors with the vectors $\bq_r$ describing the
forms. Therefore we define a basis with the following properties
\bea
 <\boe_a,\boe_b>&=&\eta_a\d_{ab} \nn\\
 <\boe_a,\boe_r>&=&0 \\
 \boe_r &=& \frac{\bq_r}{q} \; . \nn
\eea
Note that the basis vectors $\boe_r$ are no longer orthogonal to each other.
We remind the reader that the different types of indices were
defined as $i,j,...=0,...,n$ for the whole set of basis vectors,
$r,s,...=1,...,m$ for the basis vectors corresponding to forms and
$a,b,...=0,m+1,...,n$ for the other basis vectors.
With the above properties of $\boe_i$, insertion of the
expansion~\refs{expansion} into the Lagrangian~\refs{lag} leads to
\bea
 {\cal L}_0 &=& \frac{1}{2}\sum_{a}\eta_{\r_a'}^2
                +\frac{1}{2q}\sum_{r,s}A_{rs}\r_r'\r_s'-U \\
 U &=& \frac{1}{2}\sum_{r}u_r^2\exp\left(\sum_{s}A_{rs}\r_s\right)
 \; .
\eea
Recall, that $A_{rs}$ defined in eq.~\refs{Cartan} is assumed to be the
Cartan matrix of a semi--simple Lie group $G$. From the above Lagrangian, we
see that by introducing the basis $\boe_i$ we have achieved a separation
between the trivial modes which do not occur in the potential $U$ and
the nontrivial ones in the directions of the vectors $\bq_r$.

The solution of the equations of motion to be derived from ${\cal L}_0$
can be written in the form 
\be
 \ba{lll}
 \r_a &=& k_a(\t -\t_a) \\
 \r_r &=& -2 {A_{rr}}^{-1} \ln (g_r) 
 \ea \label{rho2}
\ee
with $g_r$ given by
\be
 g_r = \sum_{\bla\in\L_r}b_r(\bla )\exp (\bla \cdot\bk\t -\bla \cdot\bt )\; .
\ee
Here $\L_r$ are the weights of the fundamental representations of the
group $G$, $\bk = (k_r)$ is a constant vector restricted to the
Weyl chamber of $G$ and the set of time shifts $\bt = (\t_r )$ is arbitrary.
As before the constants $k_r$ are subject to the Hamiltonian constraint.
The coefficients $b_r(\bla )$ are determined in terms of $k_r$ and
properties of the group $G$ but the general expression~\cite{kostant}
is not very helpful. In practice, one may just insert the above solution
into
\be
 \r_r''+\frac{q}{2}u_r^2\exp\left(\sum_{s}A_{rs}\r_s\right) = 0
 \label{ugly}
\ee
to find $b_r(\bla )$.

\vspace{0.4cm}

Let us proceed with the simplest nontrivial example
$G=SU(3)$, that is, with a model having two forms excited 
and where the matrix $A$ takes the form~\cite{slansky}
\be
 A = \left(\ba{cc} 2&-1\\-1&2 \ea\right)\; .
\ee
How can such a model be realized within $D=10$ type II theories? We choose
an elementary IIA 3--form with vector $\bq_1$ and an elementary IIA
NS 2--form with vector $\bq_2$ such that they do not overlap in the spatial
directions. In the language of eq.~\refs{el_el} we then have $\d_1=3$,
$\d_2=2$ and $\d_{12}=0$ which results in $<\bq_1,\bq_1>=<\bq_2,\bq_2>=4$
and $<\bq_1,\bq_2>=-2$. The relation~\refs{Cartan} is therefore indeed
fulfilled for the above $SU(3)$ Cartan matrix if we choose $q=2$.

\vspace{0.4cm}

The weight systems of the two fundamental $SU(3)$ representations are
$\L_1 = \{ (1,0),(-1,1),(0,-1)\}$ for ${\bf 3}$ and
$\L_2 = \{ (0,1),(1,-1),$ $(-1,0)\}$ for ${\bf\bar{3}}$. The constants
$\bk = (k_1,k_2)$ are constrained by $2k_1-k_2>0$, $2k_2-k_1>0$ ($\bk$ is
in the Weyl chamber) which ensures that the argument $g_r$ of the logarithm
in eq.~\refs{rho2} is positive.
{}From eq.~\refs{ugly} we get for the coefficients $b_r(\bla )$
\be
 \ba{lllllll}
  b_1(1,0) &=& qu_1^2\,\frac{2k_2-k_1}{P}&
  &b_2(0,1) &=&q u_1^2\,\frac{2k_1-k_2}{P}\\
  b_1(-1,1) &=&q u_1^2\,\frac{k_1-k_2}{P}&
  &b_2(1,-1) &=&qu_2^2\,\frac{k_1-k_2}{P}\\
  b_1(0,-1) &=& qu_2^2\,\frac{2k_1-k_2}{P}&
  &b_2(-1,0) &=& qu_2^2\,\frac{2k_2-k_1}{P}
 \ea
\ee
with $P=(2k_2-k_1)(k_1-k_2)(2k_1-k_2)$.
The Hamiltonian constraint ${\cal H}_0=0$ in eq.~\refs{lag} turns into
\be
 {\cal H}_0 = \frac{1}{2}\sum_{a}\eta_ak_a^2+\frac{1}{2}
              (k_1^2-k_1k_2+k_2^2) = 0\; .
\ee
by inserting the complete solution.

Though significantly more complicated, the structure of these solutions
is similar to what we found in the case of orthogonal vectors $\bq$.
For $\t\rightarrow\pm\infty$ we have two asymptotic Kaluza--Klein regions.
Having chosen appropriate constants $\bk$ and $\bt$ one may also have
intermediate Kaluza--Klein regions. The transition between these
regions, however, is more complicated and the effect of the two forms
can no longer be separated in time.


\section{Including curvature}

Thus far, we have assumed that the maximally symmetric subspaces into which
we have split the total space are flat. In this section, we would like to show
that the more general case of curved subspaces can easily be incorporated into
our general framework. This is of particular interest because it
allows us to investigate a class of cosmologies, directly related to 
black $p$-brane solutions, some examples of which have
recently been given by Behrndt and F\"orste
\cite{be_fo} and Poppe and Schwager \cite{rudi}. 

Suppose, that instead of~\refs{metric} we start with a metric
\be
 ds^2 = -N^2(\t )d\t^2+\sum_{i=0}^{n-1}e^{2\a_i}d\O_{K_i}^2\; ,
 \label{c_metric}
\ee 
where $d\O_{K_i}^2$ is the metric of a $d_i$ dimensional space with constant
curvature $K_i=-1,0$ or $+1$. The effect of these terms is to create
an additional potential $V_c$ for the scale factors
$\a_i$. The effective Lagrangian~\refs{lagrangian} is
modified to 
\be
 {\cal L}_c = {\cal L}-N^2 EV_c\; ,
\ee
where $V_c$ is explicitly given by
\be
 V_c = -\sum_i 2K_ie^{-2\a_i}\; . \label{V_c}
\ee
This potential is formally very similar to the one provided
by the forms (see the eqs.~\refs{Ves}). It leads to
terms of the form $u^2\exp (\bq\cdot\bal )$ in the effective potential $U$
which, in fact, we have already considered by using the general
form~\refs{U} of $U$. All we have to do is to specify the vector
$\bq_k$ characterizing a curvature in the $k$th subspace. From eq.~\refs{V_c}
it follows that
\be 
 \bq_k = (2d_0,...,2(d_k-1),...,2d_{n-1},0)\; . \label{curv}
\ee
Moreover, the constant $u_k^2$ in front of such a curvature term in $U$ is
given by $u_k^2=-2K_k$ so that it can be of either sign depending on the sign
of the curvature $K_k$. This is in contrast to terms in $U$ describing forms
which are always positive.

All methods presented in this paper for a potential $U$ describing forms only
can now be applied to the extended version including the
curvature terms. The relevant information is encoded in the vectors
$\bq$ which we have specified for forms in eq.~\refs{elementary},
\refs{solitonic} and for curvature terms in eq.~\refs{curv} above.

For example, as in the case of flat subspaces, a model is soluble if it
corresponds to a Toda model. To check this, one has to compute the scalar
products between all the vectors $\bq$ now also including those describing
curvature in analogy to eq.~\refs{el_el}. This is easily done given the
above expression~\refs{curv}. For instance, the scalar product between a
vector $\bq_k$ describing curvature in the subspace $k$ and a vector
$\bq^{\rm (el)}$ for an elementary form is given by
\be
 <\bq_k,\bq^{\rm (el)}> = -2\e_k\; ,
\ee
where we have used eq.~\refs{s_prod}, \refs{Gin} and eq.~\refs{elementary}. 
Recall, that $\e_k=1$ if the subspace $k$ is occupied by the form and
$\e_k=0$ otherwise. Therefore, a model with one elementary form and
one curved transverse subspace has orthogonal $\bq$ vectors and can be
solved following the lines of section 4.1.
More generally, if a Toda model with curvature has been found by calculating
all scalar products, it can be solved following the general procedure
which we have described in section 4.

By way of a specific example with a non-flat subspace, let us consider
the cosmology with non-trivial NS three-form given by
Behrndt and F\"orste \cite{be_fo}. Note that, unlike all previous
examples, this is in five not ten dimensions. The solution is derived
directly from the five-dimensional black hole solution with a NS
three-form \cite{blackp} by going to a regime where the time and radial
coordinates exchange roles. The metric is given by 
\be
 ds^2 = - \frac{dt^2}{\left(-1+t_+^2/t^2\right)\left(1-t_-^2/t^2\right)} 
        + \frac{-1+t_+^2/t^2}{1-t_-^2/t^2} dy^2 + t^2 d\O^2
 \label{bf_metric}
\ee
where $d\O^2$ is the metric on a three-sphere, while the dilaton and the NS
form are given by
\be 
 \f = -\frac{1}{2} \ln \left(1-t_-^2/t^2\right) + \f_0
 \qquad
 H_{{\m_1}{\m_2}{\m_3}} = 2 t_+ t_- t^{-6} \e_{{\m_1}{\m_2}{\m_3}}
\ee
where $\m_1,\m_2$ and $\m_3$ are coordinates on the three-sphere. We see
immediately that this solution is in the form of our general
Ansatz. Comparing with \refs{c_metric}, we see that the space has been
split into two subspaces: a one-dimensional subspace parameterized by
$y$ and a three-dimensional spherical subspace which together with the
time direction constitutes our
observed spacetime. Thus we would write $d_0=1$ and $d_1=3$ while 
$K_0=0$ and $K_1=1$. The NS form field spans only the three-sphere,
without a component in the time direction. This implies a solitonic
Ansatz for $H_{{\m_1}{\m_2}{\m_3}}$. Using eqs. \refs{q_el} and
\refs{curv}, we can immediately write down the vectors $\bq_H$ and
$\bq_K$ corresponding to the NS field and the non-zero subspace curvature,
\be
 \bq_H = (2,0,8/7) \qquad \bq_K = (2,4,0) \; .
\ee
If $u_H$ is the charge of the NS field, then, by the discussion above,
the potential $U$ for this Ansatz is given by
\be
  U = \frac{1}{2} u_H^2 e^{\bq_H\cdot\bal} - 2 e^{\bq_K\cdot\bal} \; .
\ee
Similarly, we can calculate the metric on moduli space and its inverse
for this five-dimensional Ansatz using eqs. \refs{G} and
\refs{Gin}. For the inverse metric we find
\be
 G^{-1} = \left(\ba{ccc} \frac{1}{3}&-\frac{1}{6}&0\\
                         -\frac{1}{6}&0&0\\
                         0&0&\frac{3}{8}\ea\right) \; . 
\ee
This allows us to compute the scalar product between $\bq_H$ and
$\bq_K$. We find that 
\be
 <\bq_H,\bq_K> = 0 \; ,
\ee
so that, in fact, we have a model with orthogonal $\bq$ vectors. As a
result, it corresponds to a simple SU(2) Toda model and we can
solve the equations of motion exactly following the discussion given
in section 4.1. Choosing the correct time parameterization, we
thus re-derive the solution of Behrndt and F\"orste given above. We
note that there are many other models with a spherical subspace which
could be solved in this way. It would be interesting to know if these
can all be directly related to black $p$-brane solutions.

It is also interesting to note that the presence of a curved subspace
significantly alters the form of the solution. We note that, unlike the
previous cases, with a spherical subspace the corresponding term in
the potential $U$ is negative. The effect, in a solution such as that of
Behrndt and F\"orste, is that there is no longer necessarily either an
initial or a final singularity. From the form of the metric in 
eq. \refs{bf_metric}, it would appear that the solution is divergent at
$t=t_+$ and $t=t_-$. However, in fact, the scalar curvature is finite
at each point (essentially since they correspond to the two horizons
of the origin black $p$-brane solution). This raises the very interesting
possibility that solutions of this form may resolve the graceful exit
problem inherent in the cosmological solutions given thus far,
providing an inflating solution which does not end in a curvature
singularity. 


\section{Conclusion}

Throughout the paper we have noted the close relation between the
cosmological solutions we have been discussing and supergravity $p$-brane
solutions. The relationship consists of exchanging the time coordinate
for a transverse spatial coordinate, thus transforming a cosmological
space where everything evolves in time into a brane world-volume where
the fields all depend on the transverse radial coordinate, and vice
versa. This relationship is very suggestive. In particular, it raises
the possibility that the statistical mechanics of Hawking radiation in
cosmological spacetimes (which is responsible for large scale
structure) could have a similar string theoretical origin as the
entropy of black holes \cite{black_hole} (which can be computed from
the associated D-branes). We are studying this possibility and will
report on it elsewhere. 

\vspace{0.8cm}

{\bf Note:} Shortly after this paper was finished, an interesting paper by
Larsen and Wilczek~\cite{la_wi}
appeared which emphasizes that the solutions of Behrndt et. al.~\cite{be_fo},
Poppe et. al.~\cite{rudi} and some of those discussed in this paper
actually exactly correspond to the interior of certain p-brane black holes,
and gave several new cosmological solutions in this context. They
independently point out that these solutions might evade the problem of
curvature singularities.

\vspace{0.8cm}

{\bf Acknowledgments:} We would like to thank F. Larsen and F. Wilczek for
discussions. A.~L.~is supported by a fellowship from Deutsche
Forschungsgemeinschaft (DFG). A.~L.~and B.~A.~O.~are supported in part by
DOE under contract No. DE-AC02-76-ER-03071. D.~W.~is supported in part by
DOE under contract No. DE-FG02-91ER40671.
\end{document}